\RequirePackage{xcolor}

\documentclass[journal,twoside,web]{ieeecolor}

\usepackage{tmi}
\usepackage{cite}
\usepackage{amsmath,amssymb,amsfonts}
\usepackage{algorithmic}
\usepackage{graphicx}
\usepackage{textcomp}
\usepackage{orcidlink}
\usepackage{multirow}
\usepackage{caption}
\usepackage{booktabs}
\usepackage{adjustbox}
\usepackage{array}
\usepackage{hyperref}
\usepackage{nicematrix}
\usepackage{makecell}
\usepackage{generic}

\usepackage{tabularx}
\usepackage{subcaption}
\usepackage{placeins}

\usepackage{ctable} % for \specialrule command

\usepackage{color, colortbl}

\usepackage[dvipsnames]{xcolor}
\definecolor{IBMBlue}{HTML}{0F62FE}
\colorlet{MyCol}{IBMBlue!8}
\newcommand{\colrow}{\rowcolor{MyCol}}

\def\BibTeX{{\rm B\kern-.05em{\sc i\kern-.025em b}\kern-.08em
    T\kern-.1667em\lower.7ex\hbox{E}\kern-.125emX}}
\begin{document}

\title{Reinforcement Learning for Unsupervised Domain Adaptation in Spatio-Temporal Echocardiography Segmentation}

\author{Arnaud~Judge, Nicolas~Duchateau, Thierry~Judge, Roman~A.~Sandler, Joseph~Z.~Sokol, Christian~Desrosiers, Olivier~Bernard*, and~Pierre-Marc~Jodoin*\vspace{-0.3cm}
\thanks{This work was supported in part by the Fonds de recherche du Québec en Nature et Technologies ({\tiny \url{https://doi.org/10.69777/368951}}), the French National Research Agency (LABEX PRIMES [ANR-11-LABX-0063], and ORCHID [ANR-22-CE45-0029-01] project), and %within the program "Investissements d’Avenir" [ANR-11-IDEX-0007], 
the iCardio-MITACS acceleration [IT45281]). For the purpose of open access, the authors have applied a CC BY public copyright license to any Author Accepted Manuscript (AAM) version arising from this submission.\\ * Equal contribution}
\thanks{A. Judge, T. Judge and P.-M. Jodoin are with the Department of Computer Science, University of Sherbrooke, Sherbrooke, QC, Canada (e-mail: arnaud.judge@usherbrooke.ca).}
\thanks{T. Judge, N. Duchateau and O. Bernard are with  INSA, Universite Claude Bernard Lyon 1, CNRS UMR 5220, Inserm U1206, CREATIS, Villeurbanne, France.}
\thanks{C. Desrosiers is with the Dep. of Software and Information Technology Engineering, École de technologie supérieure, Montreal, Canada}
\thanks{N. Duchateau and O. Bernard are  with the Institut Universitaire de France (IUF)}
\thanks{R.A. Sandler and J.Z. Sokol are with iCardio.ai, Los Angeles, USA}
}
\maketitle
\begin{abstract}

Domain adaptation methods aim to bridge the gap between datasets by enabling knowledge transfer across domains, reducing the need for additional expert annotations. However, many approaches struggle with reliability in the target domain, an issue particularly critical in medical image segmentation, where accuracy and anatomical validity are essential. This challenge is further exacerbated in spatio-temporal data, where the lack of temporal consistency can significantly degrade segmentation quality, and particularly in echocardiography, where the presence of artifacts and noise can further hinder segmentation performance.
To address these issues, we present RL4Seg3D, an unsupervised domain adaptation framework for 2D + time echocardiography segmentation.
RL4Seg3D integrates novel reward functions and a fusion scheme to enhance key landmark precision in its segmentations while processing full-sized input videos. By leveraging reinforcement learning for image segmentation, our approach improves accuracy, anatomical validity, and temporal consistency while also providing, as a beneficial side effect, a robust uncertainty estimator, which can be used at test time to further enhance segmentation performance. We demonstrate the effectiveness of our framework on over 30,000 echocardiographic videos, showing that it outperforms standard domain adaptation techniques  without the need for any labels on the target domain. Code is available at \url{https://github.com/arnaudjudge/RL4Seg3D}.

\end{abstract}


\section{Introduction}
\label{sec:introduction}

Although supervised deep learning has become the staple for both 2D and 3D medical image segmentation, it remains limited by the amount and quality of manual annotations. Obtaining such annotations is laborious, logistically challenging, and expensive, in particular for 3D images or 2D+t image sequences. This, with accurate segmentations needed to compute clinically relevant indices such as chamber volumes or ejection fraction, has driven the development of semi-supervised and unsupervised domain adaptation methods to leverage larger datasets containing few or no annotations~\cite{guan2022}. 

Reinforcement learning (RL) offers an alternative to conventional supervised training by leveraging automated reward mechanisms to iteratively improve model outputs. We recently proposed a RL-based segmentation strategy (RL4Seg) \cite{Judge_RL4Seg} framing 2D segmentation as a single-timestep RL task, in which a segmentation network acts as an agent, and is optimized through reward-driven interactions with unlabeled data. This approach enables learning with non-differentiable objectives, including anatomical metrics, ensuring viability of output segmentations, and providing a reliable uncertainty estimator via the fully trained reward network.

However, when compared to 2D segmentation, the segmentation of 2D+t image sequences introduces an additional level of complexity to the task. The spatio-temporal consistency of segmenting
underlying structures, with variable movements, speeds and visibility, is highly challenging, especially in echocardiography, where artifacts and speckle decorrelation inherent to ultrasound further complicate the task. Models relying on 2D convolution operations independently calculated for each frame struggle with temporal consistency and smoothness, making them unsuitable for clinical use~\cite{lin2024}.
Post-processing can mitigate these issues \cite{Painchaud_2022}, yet it remains limited as it provides no guarantee that the resulting segmentations adhere closely to the ground truth \cite{Painchaud_2020}. In addition, models such as the original RL4Seg, which operate on heavily downsampled inputs (e.g., 256×256) at specific clinically relevant instants (end-diastole and end-systole), suffer from compacted information and limited temporal context, further reducing their clinical applicability.

\subsubsection*{Contributions}

In this paper, we present RL4Seg3D, an unsupervised domain adaptation framework for 3D spatio-temporal segmentation, applied to 2D + time echocardiographic sequences and targeting both the left ventricle and myocardium.
Specifically, relative to our previous work ~\cite{Judge_RL4Seg}, we make the following contributions:
\begin{itemize}
    \item We expand the segmentation RL formalism to support multiple simultaneous rewards aimed at jointly improving the policy for both general segmentation quality and task-specific constraints.
    \item We address temporal consistency across frames by enabling coherent processing of full sized input videos and designing a new reward template for temporal consistency.
    \item We introduce a landmark-based reward to explicitly constrain and improve anatomical landmark localization within spatio-temporal segmentations.
    \item We extend the uncertainty estimation capabilities of the reward network to enable pixel-wise confidence evaluation across spatio-temporal segmentations.
    \item We add a test-time optimization mechanism that leverages uncertainty estimates to improve performance on challenging videos.
    \item We establish state-of-the-art results, improving segmentation accuracy, anatomical validity, and temporal coherence over standard domain adaptation methods and foundation models, as measured by echocardiography-specific metrics, without requiring any annotations in the target domain.
\end{itemize}

\section{Previous Works}
\label{sec:previous_work}
We consider various approaches relevant to echocardiographic segmentation in the context of domain adaptation.

\textbf{\em Unsupervised methods} 
leverage unlabeled data through representation learning or direct adaptation. Pre-training strategies aim to capture relevant domain features before downstream fine-tuning. They include masked reconstruction \cite{maani2024simlvseg}, where a model predicts missing frames in a masked image sequence, and contrastive learning \cite{lamoureux2023SSL}, which aligns features across related images to learn robust representations. In contrast, pseudo-labeling \cite{ferreira2025self,Sheikh2022UDAS,Li-pseudolabels-review} offers an alternative by iteratively refining predictions using past outputs as labels, encompassing strategies like self-learning \cite{triguero2015self}, student–teacher architectures \cite{yu2019uncertainty,shen2023cotraining}, and confidence thresholding \cite{Ghamsarian2023TSIT}. Such adaptation approaches enhance feature alignment with the target domain and have been shown to improve segmentation performance.

\textbf{\em Foundation models}, namely models based on the Segment Anything (SAM) architecture~\cite{kirillov2023segment} and its variants in medical imaging~\cite{ma2024segment, zhang2023customized, wu2025medical}, can compensate for domain shifts by leveraging strong generalization capabilities learned through training on large heterogeneous datasets.

Although SAM-based methods perform well across modalities, video segmentation remains challenging due to temporal inconsistency in frame-by-frame processing and the impracticality of per-frame prompting. Recent approaches address this by integrating tracking mechanisms into the SAM architecture~\cite{cheng2023samtrack,yang2023trackanything}.

For ultrasound imaging, models such as SAMUS \cite{lin2024samus} have been trained on large ultrasound datasets. Building on this, MemSAM \cite{deng2024memsam}, a video adaptation of SAMUS for echocardiography, has improved performance on 2D + time sequences. It does so by incorporating specifically designed memory modules to automatically generate prompts and reduce the propagation of the noise inherent to ultrasound images.

\textbf{\em Image-to-image translation} generates plausible target-domain images from labeled source-domain data, enabling supervised training on target-style images with source labels. Approaches include diffusion models \cite{azuma2025zodi} and generative adversarial networks (GANs) \cite{skandarani2023gans,zhu2017unpaired,CHEN2022_GAN_medai }
with applications
including ultrasound domain adaptation \cite{iacono2024structurecyclegan,HUANG2022uscyclegan}.

Unfortunately, these approaches face important limitations. Synthetic translations may distort anatomical structures, introduce artifacts, or retain residual source-domain features, leading to label–image mismatches and reduced clinical reliability~\cite{skandarani2023gans}. Moreover, when labeled source data is scarce and the target domain is large and heterogeneous, these methods often fail to capture the full anatomical variability and subtlety of target images, limiting their adaptation effectiveness.

\textbf{\em Reinforcement learning (RL)} 
involves methods that enable an autonomous agent to learn from interactions with its environment without the need for a differentiable loss function nor any annotated data~\cite{jaeger2024invitation}. During training, the agent observes states, chooses an action based on its current policy and transitions to a subsequent state, receiving a reward reflecting the quality of the action within that state. This feedback allows for iterative improvement of the policy over time.

RL has been used to align the model output with human preferences with an approach called Reinforcement Learning from Human Feedback (RLHF) \cite{kaufmann2023survey, christiano2017_rlhf_gym}.
This method has proven highly effective for ensuring high-quality responses from large language models \cite{ziegler2019fine, ouyang2022training, stiennon2020learning}, including ChatGPT.

In the context of medical imaging, RL is generally used in refinement or adjustment of predictions for tasks such as landmark prediction \cite{ALANSARY2019156, Ghesu2019}.
More specifically in image segmentation, it is often limited to proxy tasks such as hyperparameter search, active learning and region of interest detection \cite{hu2023reinforcement}, with a few exceptions explicitly targeting segmentation \cite{xu2025rlcoseg}. One exception is RL4Seg \cite{Judge_RL4Seg}, which enables direct end-to-end domain adaptation segmentation by formulating the task as a single-timestep trajectory. The input image is the only state in the trajectory, the action is a predicted segmentation map, and the reward is a pixel-wise error map of the segmentation. The reward is learned from a reward dataset of  valid and invalid segmentations of the same image, and is used to provide feedback to improve the segmentation policy via the proximal policy optimization (PPO) algorithm \cite{schulman2017proximal}, following an approach inspired by RLHF.
By relying on flexible reward driven feedback rather than target domain annotations, reinforcement learning can enforce task-specific priors such as the anatomical metrics on the policy. This improves segmentation quality across domains and enables the approach to be applied to other tasks or imaging modalities wherever such validity constraints can be defined.

\section{Method}
\label{sec:method}

Our unsupervised domain adaptation framework for 3D segmentation starts with a limited set of $n$ pairs of labeled data from a source domain, denoted as \mbox{$\mathcal{D}_S = \{(\mathbf{x}^{(1)}_S, \mathbf{y}^{(1)}_S), ..., (\mathbf{x}^{(n)}_S, \mathbf{y}^{(n)}_S) \}$} where $\mathbf{x}$ is an 2D+t image sequence and $\mathbf{y}$ is the corresponding 2D+t segmentation map. Through supervised pre-training on this source data, the segmentation network learns a strong prior, which serves as a foundation for the subsequent reinforcement learning-based domain adaptation. This adaptation process aims to refine the segmentation network to ensure it produces accurate, anatomically valid and temporally consistent segmentations on a large-scale target domain \mbox{$\mathcal{D}_\mathcal{T} = \{\mathbf{x}^{(1)}_\mathcal{T}, ..., \mathbf{x}^{(m)}_\mathcal{T}\}$}, which consists exclusively of $m$ unlabeled image sequences.

To achieve this, a novel reward fusion scheme merges complementary feedback from multiple sources to drive the optimization of the pre-trained segmentation network with reinforcement learning. This approach incorporates both adaptive rewards that are continuously refined during training and static rewards that are pre-trained and fixed during the domain adaptation process. Leveraging both domain and task-specific priors with adaptive feedback enables balanced and flexible guidance to improve segmentation performance on the target domain.

\begin{figure*}[tp]
    \centering \includegraphics[width=1\linewidth]{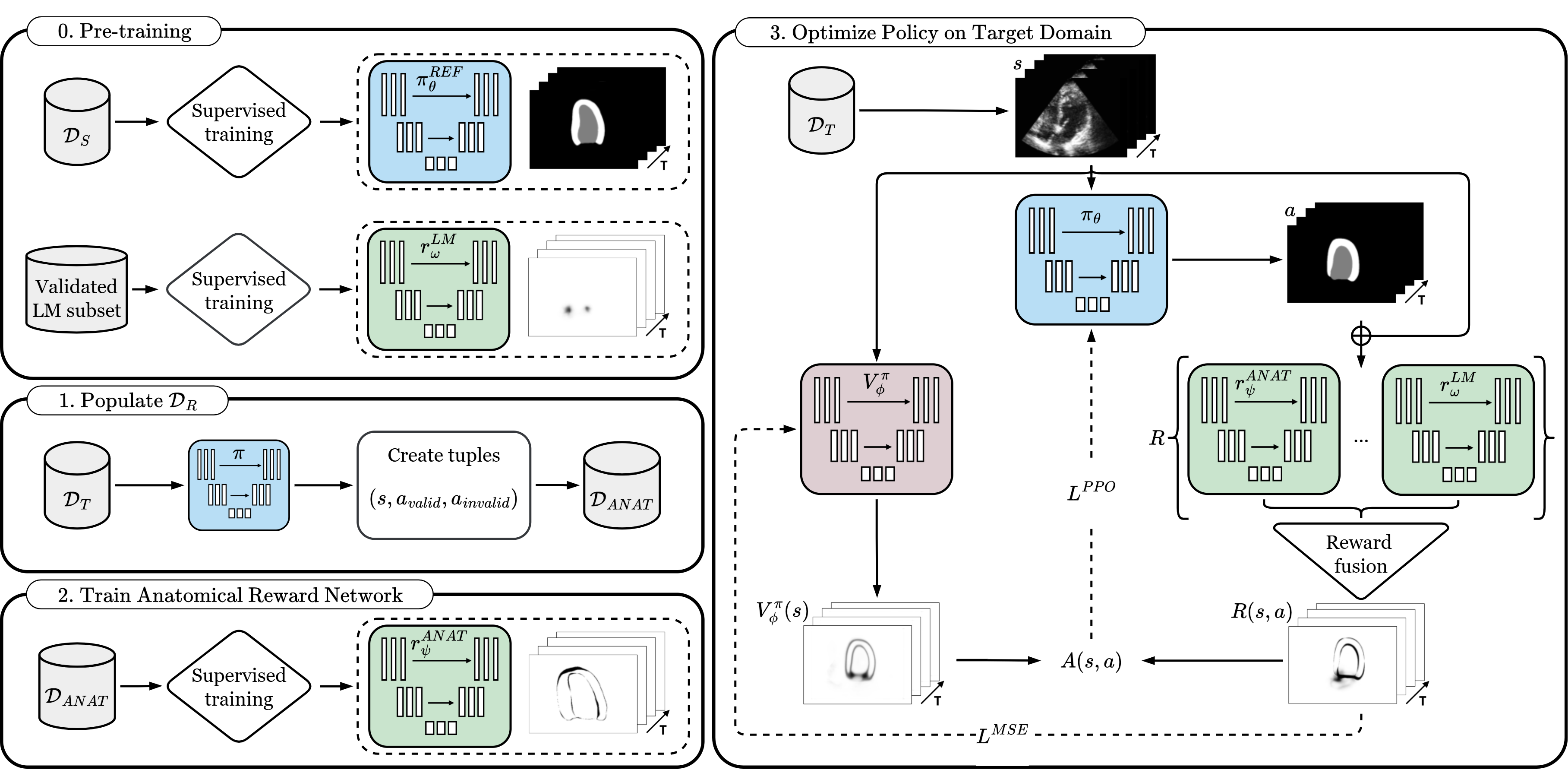}
    \caption{Overview of the full RL4Seg3D framework and its main steps.  ${\cal D}_s$: labeled source domain dataset, ${\mathcal D}_\mathcal{T}$: unlabeled target domain dataset, ${\cal D}_{\mathit{ANAT}}$: anatomical reward dataset, $\pi_\theta$: policy, $s$: state, $a$: action, $R(s,a):$ reward function,  $V_\phi^\pi(s):$ value function, $r_{\psi}^{ANAT}$: anatomical reward network, $r_{\omega}^{LM}$: landmark reward network, $A(s,a)$: advantage function.
    \vspace{-0.5cm}}
    \label{fig:overview}
\end{figure*}

\subsection{3D Segmentation RL}

As for its 2D counterpart \cite{Judge_RL4Seg}, 3D segmentation RL operates on the basis of single timestep trajectories. States $s$ are defined as a temporal patch of a 2D+t image, a subsequence of consecutive full-sized frames, and actions $a$ are defined as the corresponding consecutive segmentation maps.
Although the task of spatio-temporal segmentation inherently contains a sequence of images, this series of images is considered a single state. In this way, leveraging 3D convolution improves temporal consistency, as neighboring frames are considered by the segmentation policy. This framing also allows the formalism to remain consistent for volumetric data. 
The main elements in 3D segmentation RL are defined as follows:

\subsubsection{Policy} $\pi : \mathbb{R}^{H \times W \times T} \rightarrow [0, 1]^{K\times H \times W \times T}$.
The policy is the segmentation network that is optimized for segmentation of $K$ classes on the target domain. It is modeled with a 3D U-Net which outputs a categorical distribution over each pixel of an input temporal patch from an image sequence of full height $H$ and width $W$, of length $T$. During training, actions $a \!\in\! \{0,..., K\!-\!1\}^{H \times W \times T}$ are obtained through sampling of the policy's output distribution $\pi_{\theta}(a|s)$, whereas at inference, actions are obtained deterministically via the $\arg\max$ operator.

\subsubsection{Rewards} $r(s, a): \mathbb{R}^{2 \times H \times W \times T} \!\rightarrow\! [0, 1]^{H \times W \times T}$. 
Reward functions $r(s, a)$ return a pixel-wise error map indicating high and low quality regions of a segmentation based on the concatenated segmentation and image slices. An arbitrary number of reward functions can be used and merged into the overall reward $R(s, a)$, as described in detail in Sec.\ref{sec:method_rewards}.

\subsubsection{Q function} $Q^\pi(s, a) : \mathbb{R}^{2 \times H \times W \times T} \!\rightarrow\! [0, 1]^{H \times W \times T}$. The Q function provides a measure of the expected total reward for action $a$ at state $s$, following the trajectory until its end. As there is only one state per trajectory in 3D segmentation RL, the Q-function's Bellman equation can be simplified, equating directly to the total rewards: $Q^{\pi}(s, a) = R(s, a)$.

\subsubsection{Value function} $V^\pi(s): \mathbb{R}^{H \times W \times T} \rightarrow [0, 1]^{H \times W \times T}$. The value function is a pixel-wise measure of the expected segmentation quality for a given state $s$ across all actions the current policy $\pi$ could take, with simplified Bellman equation 
$V^\pi(s) = \mathbb{E}_{a \sim \pi(\cdot | s)} [R(s, a)]$. It is modeled with a 3D U-Net, approximating the expected reward from 
merged rewards $R$.

\subsubsection{Advantage function}
The advantage function quantifies the relative quality of an action compared to the average action taken by the policy. It is defined as the difference between the Q-function and the value function: $A^\pi(s, a) = Q(s, a) - V^\pi(s)$. In 3D segmentation RL, it is simplified to $A^\pi(s, a) = R(s, a) - V^\pi(s)$.

\subsection{Training framework}

The RL4Seg3D framework can be divided into a 3-step loop and a pre-training phase, as described in Fig.\ref{fig:overview}.
\begin{enumerate}
    \addtocounter{enumi}{-1}
    \item (\emph{Pre-training}): Pre-train the segmentation policy on the source domain $\mathcal{D}_S$. This policy, $\pi_{\theta_{\mathit{REF}}}$, serves as the starting point for the following RL domain adaptation and as a frozen reference policy used to limit divergence during optimization. Static rewards are also trained or precomputed during this phase.
    \item Populate a reward dataset $\mathcal{D}_{\mathit{ANAT}}$ for anatomical correctness using the current policy $\pi$, according to the procedure described in Sec.\ref{sec:reward_ds}.
    \item Train the anatomical reward network (adaptive reward) with $\mathcal{D}_{\mathit{ANAT}}$, using binary cross-entropy.
    \item Optimize the policy $\pi$ on the target dataset $\mathcal{D}_\mathcal{T}$ against all merged rewards using the PPO algorithm. Simultaneously train the value function based on current rewards with a mean squared error loss.
\end{enumerate}

Multiple iterations of the RL loop (steps 1, 2 and 3 in Fig.~\ref{fig:overview}) allow for gradual optimization of the policy and sufficient divergence from the original reference policy. 

\subsubsection{Anatomical Reward Dataset}
\label{sec:reward_ds}
$\mathcal{D}_{\mathit{ANAT}}$.
The adaptive anatomical reward dataset contains pairs of valid and invalid 2D+t segmentations. This data can therefore be used to train an adaptive reward, in this case an anatomical reward network, using simple supervised training with binary cross-entropy loss. The ground truth target for this optimization is the pixel-wise difference between valid and invalid 2D+t segmentation masks, indicating error regions.

In order to create these pairs of segmentations, a first round of inference is done with the current policy. Anatomical and temporal metrics (c.f.~\ref{sec:eval_metrics}) are used to identify valid and invalid segmentations. When a segmentation is valid, many varied, controlled distortions are applied to the model weights and the input image sequence in order to produce multiple invalid segmentations.
These distortions include small random perturbations of the model weights, as well as Gaussian blurring and contrast reduction applied to input images.

This creates a set of invalid/valid pairs of segmentations for a given image sequence which are added to $\mathcal{D}_{\mathit{ANAT}}$. As for initially invalid segmentations, they are corrected with a variational auto-encoder (VAE) \cite{Painchaud_2022}, in order to obtain a valid segmentation. These pairs of segmentations are also added to $\mathcal{D}_{\mathit{ANAT}}$. The VAE, trained on the source domain using the anatomical and temporal metrics (cf. Sec.\ref{sec:eval_metrics}), takes only a segmentation mask as input and warps the segmentation to the closest valid representation in the latent space, while ensuring smooth temporal evolution. This promotes temporal consistency in the reward dataset and subsequently in the reward network's predictions. As the VAE does not guarantee correctness with respect to the underlying image and anatomy, only corrected segmentations exhibiting high similarity to the original invalid ones, as measured by Dice score, are retained.

\subsection{Rewards}
\label{sec:method_rewards}

\begin{figure}[ht]
    \centering
    \begin{subfigure}[t]{0.475\linewidth}
        \centering
        \includegraphics[width=\linewidth]{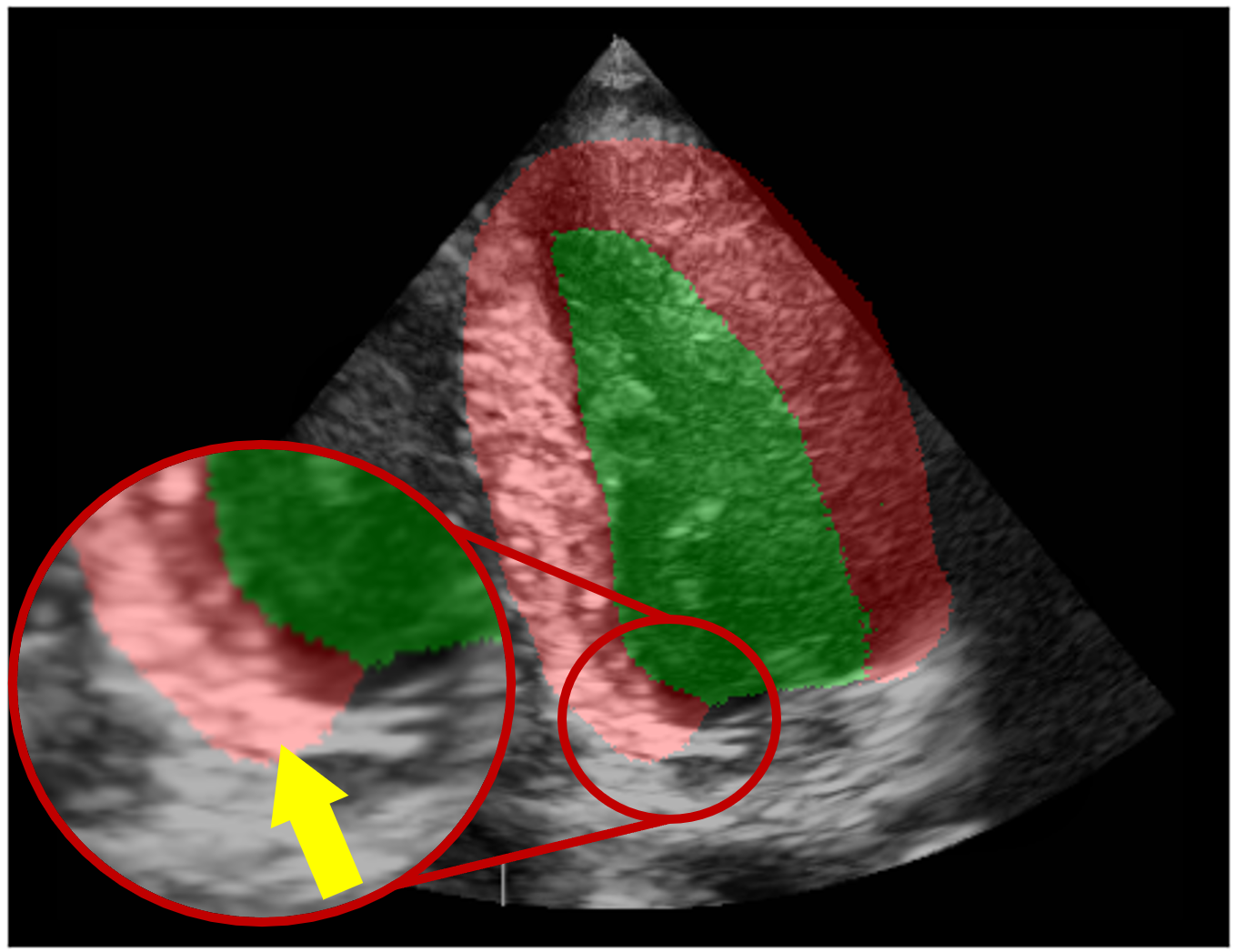}
        \caption{Segmentation map.}
        \label{subfig:r_segmentation+overlay}
    \end{subfigure}
    \begin{subfigure}[t]{0.475\linewidth}
        \centering
        \includegraphics[width=\linewidth]{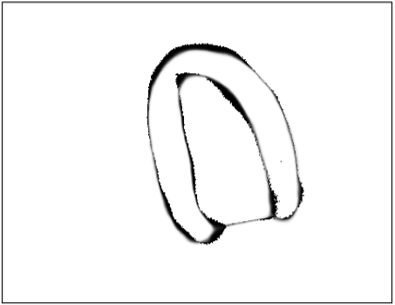}
        \caption{Anatomical reward $r^{\mathit{ANAT}}$.}
        \label{subfig:r_anat}
    \end{subfigure}
    \begin{subfigure}[t]{0.475\linewidth}
        \centering
        \includegraphics[width=\linewidth]{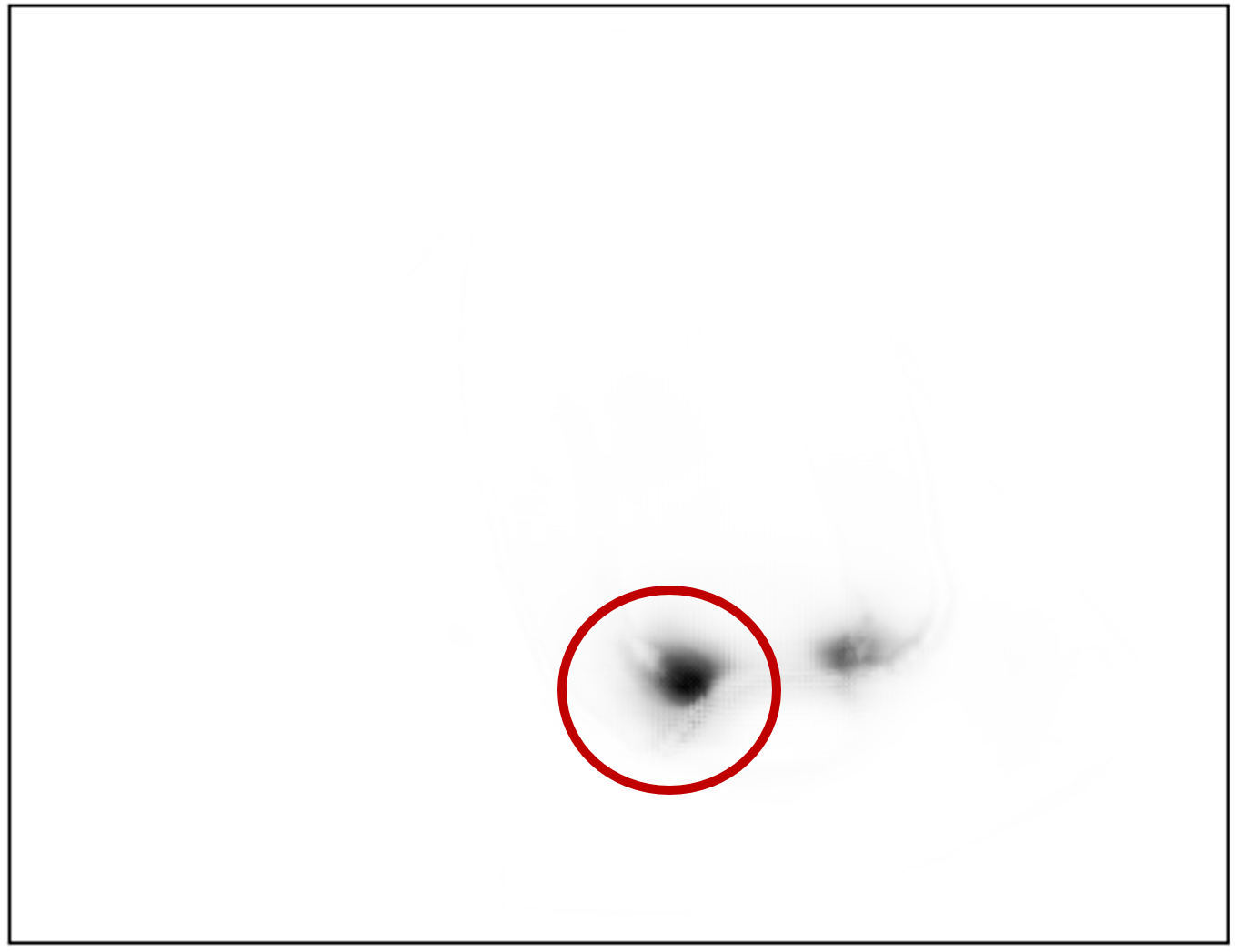}
        \caption{Landmark reward $r^{LM}$.}
        \label{subfig:r_lm}
    \end{subfigure}
    \begin{subfigure}[t]{0.475\linewidth}
        \centering
        \includegraphics[width=\linewidth]{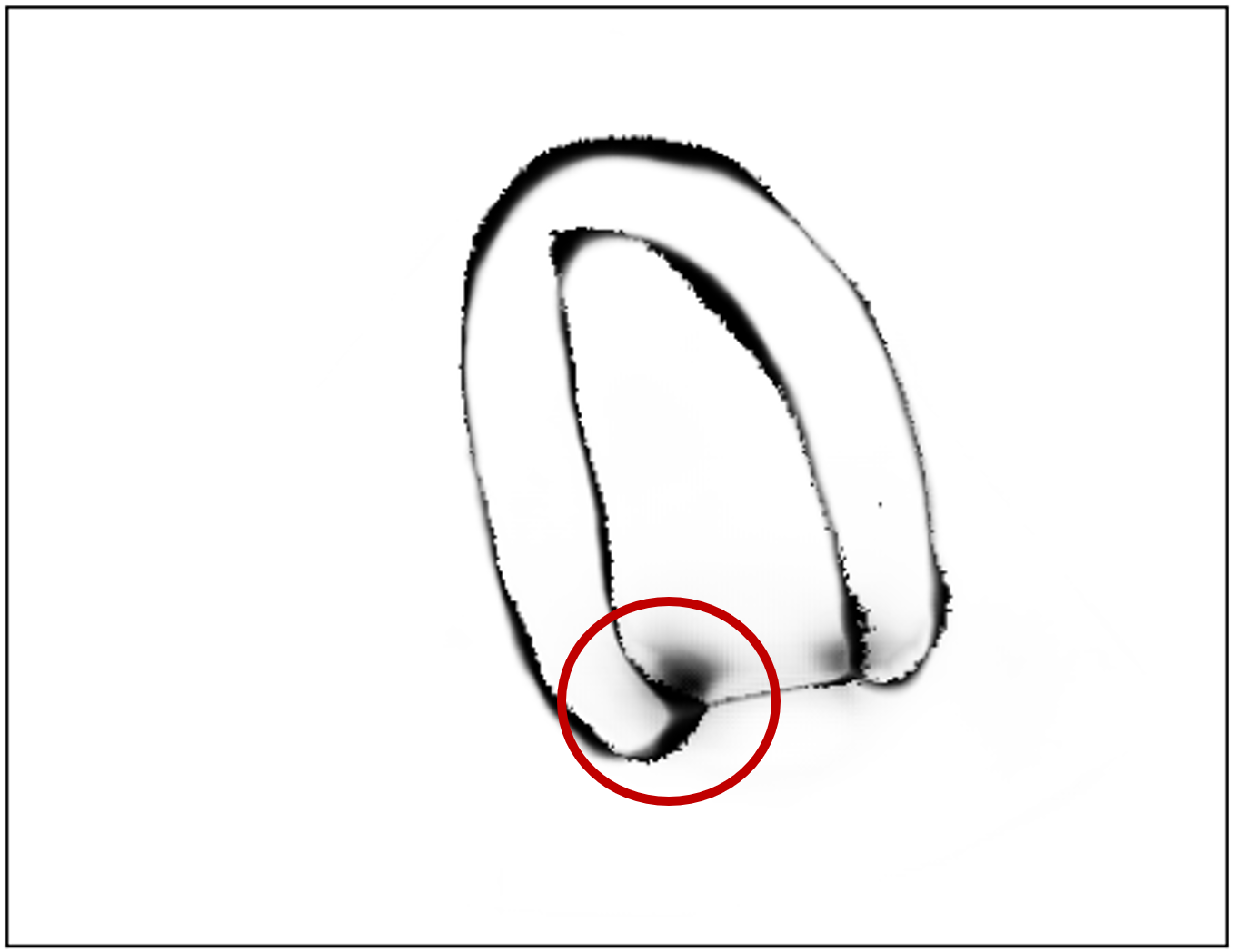}
        \caption{Merged reward map.}
        \label{subfig:r_merged}
    \end{subfigure}
    \begin{subfigure}[t]{0.475\linewidth}
        \centering
        \includegraphics[width=\linewidth]{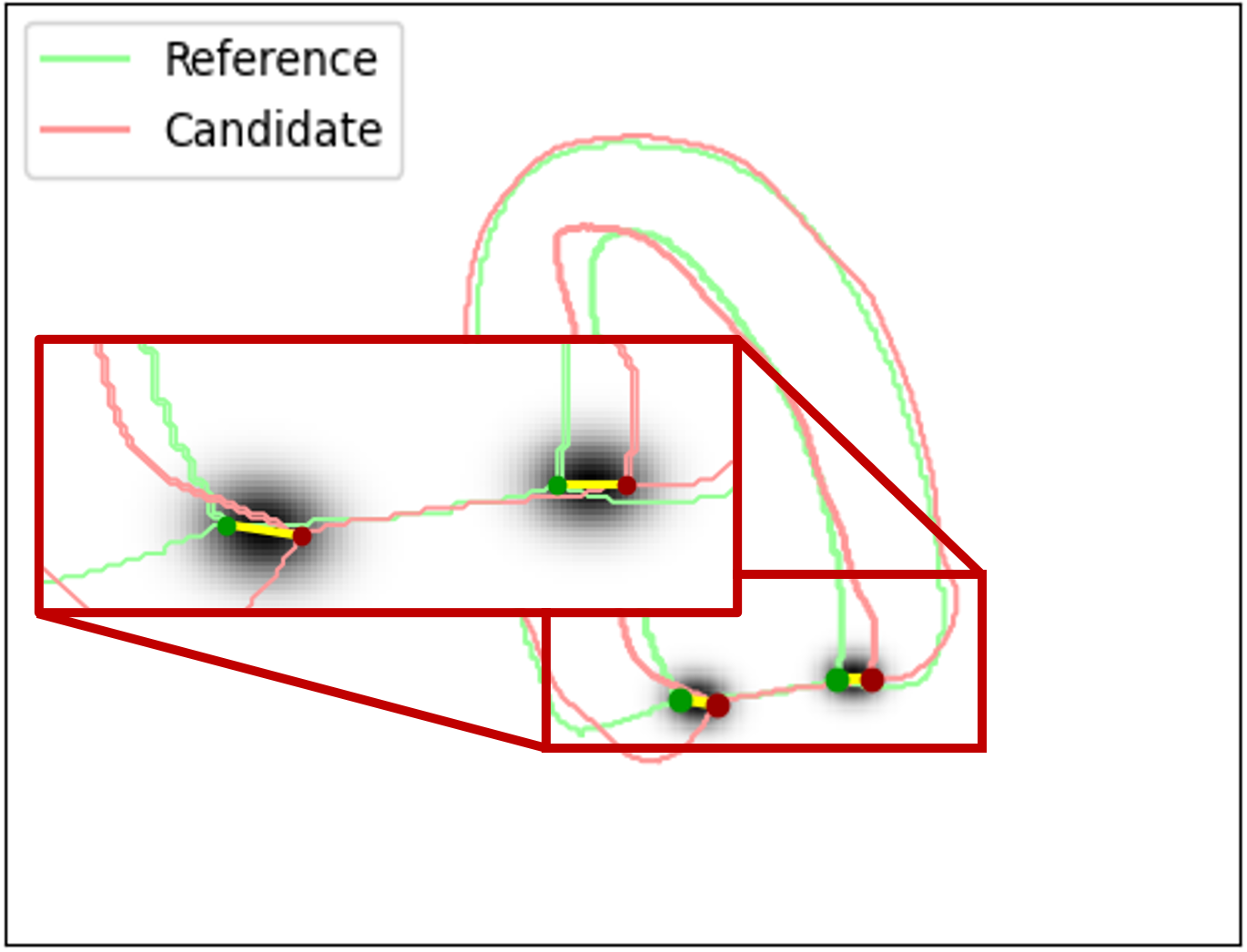}
        \caption{LM subset ground truth.}
        \label{subfig:val_LM_subset}
    \end{subfigure}
    \begin{subfigure}[t]{0.475\linewidth}
        \centering
        \includegraphics[width=\linewidth]{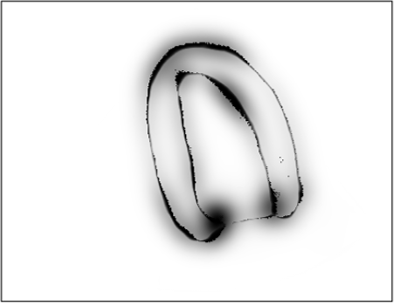}
        \caption{Effect of $P_{\mathit{Temporal}}$.}
        \label{subfig:temp_pen}
    \end{subfigure}
    \caption{Example of a segmentation frame (a) and reward maps highlighting errors. The anatomical reward (b) from $r_{\psi}^{\mathit{ANAT}}$ shows general segmentation issues such as the misshapen apex, while the landmark reward (c) from $r_{\omega}^{LM}$ highlights errors in mitral valve alignment. The final reward map (d), obtained with min-based fusion, illustrates a localization error at the left mitral valve commissure: the segmentation places it inside the ventricle (red circles), whereas the correct location is indicated by the yellow arrow. (e) illustrates ground truth error maps for the validated landmark (LM) subset, and (f) shows the fused reward map with the temporal penalty applied. Video of full sequence and rewards available in supplementary material.\vspace{-0.5cm}}
    \label{fig:main_rewards}
\end{figure}

A key strength of RL and, by extension, of the segmentation RL formalism, is its flexibility in optimizing any non-differentiable objectives. This allows multiple reward components to be incorporated into the framework to specifically address segmentation errors. Accordingly, we define a set of rewards $R$ used for 3D segmentation RL training, where each individual reward $r$ corresponds to a pixel-wise error map providing feedback on a distinct type of segmentation errors. Rewards are categorized as adaptive, which evolve and improve at each iteration of the RL framework, or static, which are pre-trained in step 0 and remain fixed throughout the domain adaptation process.

To account for errors detected by each reward, reward fusion allows for the aggregation of feedback from all rewards in $R$, through a fusion operator $f_\mathrm{fusion}$. At pixel $(i,j)$ in frame $t$, the advantage function is defined by the rewards, divergence constraint $C_{\mathit{KL}}$ (see Sec.\ref{sec:policy_opt}) and value function as:
% \begin{align}
% \label{eq:multiple_reward_adv}
%     A(s,a)_{i, j, t} = \left( \min_{r_{i,j,t} \in R_{i,j,t}} \!\!\!%\left( 
%     r_{i,j,t} - {C_{\mathit{KL}}}_{i, j, t} %\right)
%     \!\right) - {V^\pi(s)}_{i,j,t},
% \end{align}
\begin{align}
\label{eq:multiple_reward_adv}
    A(s,a)_{i, j, t} = f_{\mathrm{fusion}}(R_{i,j,t}) - {C_{\mathit{KL}}}_{i, j, t} %\right)
    \! - {V^\pi(s)}_{i,j,t},
\end{align}
where $f_\mathrm{fusion}$ can be any aggregation operator. We use the minimum operator,
\begin{align}
\label{eq:fusion_min}
    f_{\mathrm{fusion}}(R_{i,j,t}) =\min_{r_{i,j,t} \in R_{i,j,t}} \!\!\! r_{i,j,t}.
\end{align}
ensuring that the policy is corrected based on the most severe error at each pixel, maximizing its ability to address critical segmentation mistakes.

\subsubsection{Anatomical Rewards}

The anatomical reward $r_{\psi}^{\mathit{ANAT}}$ is an adaptive network based on anatomical metrics \cite{Painchaud_2020} that guides the domain adaptation process of the segmentation policy in RL4Seg3D.  As shown in figure~\ref{fig:main_rewards}, the input of this reward network is a 2D+t image patch along with its corresponding segmentation map. This network outputs a pixel-wise error map, indicating both the location and probability of anatomical errors through variable reward values assigned to each pixel. As the anatomical reward dataset $\mathcal{D}_{\mathit{ANAT}}$ is updated with new segmentation maps at each iteration of the RL framework, it becomes increasingly representative of the current policy’s typical errors, allowing the anatomical reward network to adapt to the latest segmentation behavior.

\subsubsection{Landmark Based rewards}

In spatio-temporal segmentation, a key consideration is the alignment of the output segmentation with the underlying image. A segmentation shifted by only a few pixels may still score well on metrics such as Dice coefficient or Hausdorff distance, yet remain incorrect relative to anatomical structures in the image.

To address this, we introduce a reward based on key landmarks in the image and segmentation. This reward identifies regions with probable errors related to alignment of specific structures, therefore penalizing segmentations that fail to align precisely with these landmarks during RL training. Similar to the anatomical reward network, a neural network, $r_{\omega}^{LM}$, processes an image-segmentation pair and outputs an error map
which highlights regions of low reward corresponding to misalignment of key anatomical points.

As shown in Fig.~\ref{subfig:r_lm}, we set the mitral valve commissure as the primary landmark that segmentations must accurately follow to remain consistent with the image sequence. These structures represent the junction between the mitral valve leaflets and the annulus. In apical two- and four-chamber echocardiographic views, these landmarks are generally visible on the images and correspond to the base of the segmentation mask, %located 
at the two points where the left ventricle, myocardium and background classes coincide. Aligning precisely with the mitral valve commissure is especially important for downstream tasks such as cardiac tracking, which rely on the accuracy of segmentations.

The landmark reward is designed to be robust to abnormal anatomy, such as mitral valve prolapse. In such cases, where valve leaflets may deform or flip, commissure locations remain unchanged. Any ambiguity in commissure alignment leads to a localized error, reflecting uncertainty and enabling reliable policy guidance, even in anatomically atypical cases.

The landmark reward is pre-trained using a validated landmark subset of 777 segmentation maps (from source and target domains) with manually-verified mitral valve commissure locations and is kept frozen during the iterative RL training. 
No manual ground truth annotations are used in the target domain. Source domain samples rely on the ground truth annotations used in policy pre-training, while target domain references consist exclusively of human-validated pseudo-labels.

At first, an arbitrary policy is used to generate candidate segmentations for these validated sequences. The reward network is then trained in a supervised manner, using these candidate segmentations along with their corresponding image sequences as inputs, ground truth error maps, and a binary cross-entropy loss. The error maps are created by comparing the reference and candidate segmentation's landmark points, extracted from the base of each segmentation (see Fig.\ref{subfig:val_LM_subset}). A line is drawn between each corresponding point on a blank error map, which is subsequently convolved with a Gaussian kernel of size proportional to the distance between the landmark points, creating variable sized error ellipses that highlight the erroneous regions in the candidate segmentation. This process is computed independently for each frame in the sequence, ensuring that segmentation errors are accurately captured at each time step.

\subsubsection{Temporal Reward Penalty}

Another critical aspect of spatio-temporal segmentation is temporal consistency. The use of 3D convolutions improves the temporal accuracy of the segmentation policy, however, to further reinforce consistency, we introduce a temporal penalty $P_{\mathit{Temporal}}$ as a static reward mechanism. This penalty reduces the reward values for temporally inconsistent segmentations and can be easily incorporated to the reward fusion mechanism by modifying the reward map $r$ after fusion, as needed.

Temporal validity is assessed using eight temporal metrics \cite{Painchaud_2022}, which evaluate the stability of the segmentation across frames. These metrics extract features from each frame to detect inconsistencies in temporal evolution (see Sec.\ref{sec:eval_metrics}). When an inconsistency is identified, the reward maps of the corresponding and neighboring frames are penalized by convolving them with a Gaussian kernel and re-scaling to the original intensity range. The Gaussian kernel is parametrized by $\sigma(\eta_t, \kappa)$, increasing the standard deviation according to the number of inconsistencies $\eta_t$ at time $t$ and is modulated by constant scaling factor $\kappa$,
\begin{align}
\label{eq:reward_penalty}
    P_{\mathit{Temporal}}(r)_t =
    \begin{cases}
     r_t,  & \text{if $\eta_t = 0$}, \\
     \mathit{norm} (G_{\sigma(\eta_t, \kappa)} \ast r_t), & \text{if $\eta_t > 0$}.
    \end{cases}
\end{align}

This process lowers the reward in areas surrounding already uncertain or erroneous pixels by a magnitude proportional to the number of detected inconsistencies, as these regions are most often linked to the temporal inconsistencies (see Fig.\ref{subfig:temp_pen}). Frames without inconsistencies remain unchanged, ensuring no unnecessary modifications are made. Over multiple training epochs, the policy gradually learns this temporal smoothness prior, leading to more stable and coherent segmentations.

\subsubsection{Policy Optimization}
\label{sec:policy_opt}
The Proximal Policy Optimization (PPO) algorithm is used to fine-tune the policy on the unlabeled target domain (c.f. Fig.~\ref{fig:overview}.3). Specifically, we use the clipped version of the PPO loss \cite{schulman2017proximal}, which constrains the importance sampling ratio $\rho(\theta)$, the ratio of action probabilities under the current and old policies, parametrized by different values of $\theta$, to the range $[1 -\epsilon, 1 + \epsilon]$ (with $\epsilon=0.2$). While other policy-based RL algorithms could be used, PPO provides a stable and flexible optimization process, capable of handling granular reward signals, with relative technical simplicity. Value-based approaches would be impractical due to the high-dimensional action space.

Our advantage map $A$ weights the log-probabilities of the actions predicted by the policy on a pixel-wise basis, with constraints to maintain gradual and stable optimization. The resulting objective is defined as:
\begin{equation}
    L^{\mathit{CLIP}}(\theta) = \mathbb{E}_\theta \big[\textrm{min}(\rho(\theta)A,\ \textrm{clip}(\rho(\theta);\, 1 \!-\! \epsilon, 1 \!+\! \epsilon)A)\big].
\end{equation}

As this objective is maximized, the probabilities of high reward actions are increased as those for low-reward actions are decreased. The pixel-wise nature of RL4Seg3D allows this to be granular, offering localized feedback to the policy, for specific erroneous pixels.

An entropy term based on the policy distribution is added to the loss, maintaining sufficient entropy in the distribution throughout training in order to ensure adequate exploration of the action space: $L^H = -\sum \pi_\theta \log(\pi_\theta)$.

All components of the PPO objective, including log-probabilities, importance ratios, rewards, as well as the clipping and minimum operations, are computed pixel-wise. The resulting values are then aggregated over spatial and temporal dimensions before adding the entropy term.
The full loss function that is maximized is therefore:
\begin{align}
\label{eq:full_loss}
    L^{\mathit{PPO}} = L^{\mathit{CLIP}} + \alpha L^H.
\end{align}

A divergence constraint is also introduced \cite{stiennon2020learning}, with respect to the initial reference policy, reducing the value of the rewards by a factor proportional to the KL divergence between current and reference polices:
\begin{align}
\label{eq:div_constraint}
    C_{\mathit{KL}} = \beta \cdot \mathit{KL}\big(\pi_{\theta_{\mathit{REF}}}(a|s)|| \pi_\theta(a|s)\big).
\end{align}

This allows the policy's output distribution to remain relatively close to that of the reference policy, which has learned a strong prior from source domain pre-training, preventing excessive updates that could produce a suboptimal policy.
This implies that the starting policy must be sufficiently performant to provide a meaningful reference distribution for stable optimization.
As this constraint is subtracted from the reward term (see Eq.\ref{eq:multiple_reward_adv}), it limits $L^{\mathit{CLIP}}$ by reducing the advantage term's magnitude.

\subsection{2D+t Sequences}
\label{sec:2dt_sequences}
Since the network policy and the entire 2D+t echocardiographic data are too large to fit into memory, RL4Seg3D employs a temporal sliding window.  As such, each input consists of a temporal patch comprising 4 full-size consecutive frames. During training, patches are randomly extracted from videos  to form batches, while during inference, all patches are computed sequentially and the corresponding segmentation results are merged using Gaussian averaging.

Following the approach of nnU-Net \cite{Isensee2021-nnunet}, a common pixel spacing is used to train the model in order to maintain consistent size and proportions of the underlying anatomical structures. The common spacing is determined based on the average pixel spacing in the target dataset, computed independently for each spatial axis, keeping the time axis constant.

Since full-sized time slices are used alongside a standardized pixel spacing, each sequence can have a unique image size. To handle this, all images are resampled to the common spacing and then adjusted, using minimal cropping and padding, so their spatial dimensions are multiples of the model’s stride. This ensures compatibility with the up- and down-sampling layers of the policy's 3D U-Net. During inference, inverse transformations restore images and corresponding segmentations to their original spacing and dimensions.

Due to the varying image sizes, a batch size of 1 is used during training. However, to accelerate training and leverage the benefits of mini-batch optimization, we implement distributed data parallel training using the PyTorch library. The effective batch size during training is equal to the number of GPUs.

\begin{figure*}[ht]
    \centering
    \begin{subfigure}[t]{0.45\linewidth}
        \centering
        \includegraphics[width=\linewidth]{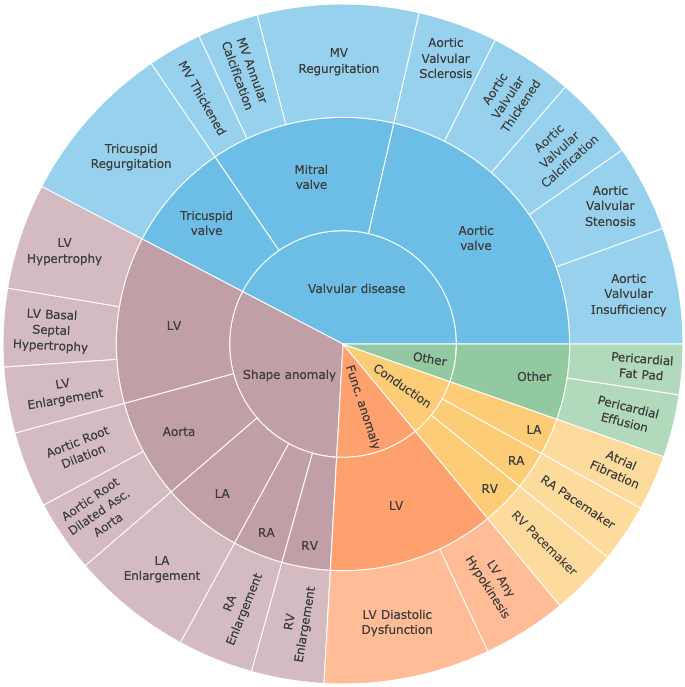}
        \caption{Entire dataset.}
        \label{subfig:patho_all}
    \end{subfigure}
    \begin{subfigure}[t]{0.45\linewidth}
        \centering
        \includegraphics[width=\linewidth]{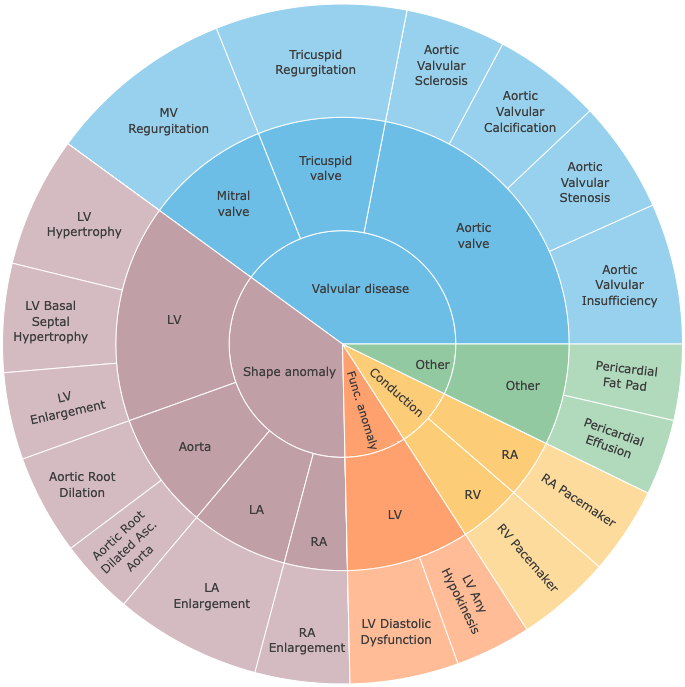}
        \caption{Test subset.}
        \label{subfig:patho_test}
    \end{subfigure}
    \caption{Distribution of pathologies present in the target dataset $\mathcal{D}_\mathcal{T}$, categorized by type and anatomical structure. \vspace{-0.5cm}}
    \label{fig:icardio_data}
\end{figure*}

\subsection{Uncertainty-Guided Test-Time Optimization}
\label{sec:post-proc-val}
\textit{Uncertainty estimation.} Unsupervised domain adaptation presents unique challenges regarding the evaluation of newly generated segmentations on the target domain. As ground truth annotations are unavailable for these images, human validation remains the most reliable approach. However, it is costly, time-consuming and subject to inter-observer variability, highlighting the need for complementary automatic segmentation evaluation metrics.

The anatomical and temporal metrics integrated into RL4Seg3D's training framework provide a strong foundation for assessing a segmentation's validity. If all frames are anatomically valid and temporal consistency is preserved throughout the sequence, there is a high likelihood that a segmentation is valid. However, these metrics evaluate only the segmentation itself, without directly considering its alignment with the underlying image, which limits their ability to detect uncertain or structurally inconsistent regions.

In RL4Seg3D, the 3D anatomical reward network provides pixel-wise uncertainty estimates that account for both the spatial structure within frames and the temporal dynamics across frames, without the need for further training following the RL loop (c.f. Fig.~\ref{fig:main_rewards}). After training, we use temperature scaling \cite{guo2017calibration} to calibrate  the model, using the validation subset to compute the optimal temperature parameter.

\textit{Test-time optimization.} The uncertainty maps can also be leveraged to refine the policy at test-time in an unsupervised manner, targeting the most challenging videos. We implement a test-time optimization (TTO) scheme, applied to videos where the policy produces segmentation containing one or more anatomical or temporal errors. As reward maps provide direct feedback on uncertain or erroneous pixels, PPO can be used to improve the policy on a sequence-specific basis. Each video is split into temporal patches, and the PPO loss, averaged over three random augmentations (Gaussian noise and contrast variation), is used to update the weights over four iterations across all patches. The augmentations promote diversity in the policy and reward outputs, improving optimization on difficult cases by encouraging reliance on strong learned priors rather than potentially ambiguous image features. To avoid degradation, the first iteration is kept frozen as a baseline, and the final weights are selected from the iteration with the highest minimum per-frame average reward. This process is applied independently for each video with weight updates discarded after producing the final prediction using the sliding window (Sec.\ref{sec:2dt_sequences}).

\section{Experiments and Results}
\subsection{Data}
Experiments were conducted on 2D+t echocardiographic sequences, with data being divided into two distinct domains. All images were preprocessed using adaptive histogram equalization to enhance contrasts, aiming to standardize image characteristics across both domains and all images.

\subsubsection{Source Domain}
The source domain $\mathcal{D}_S$ consists of 579 fully annotated echocardiography videos acquired in the apical 4-chamber (A4C) and 2-chamber (A2C) views during clinical examinations at the University Hospital of Lyon, France. The study was approved by the local ethics committee, and all participants provided written informed consent.
The videos were acquired from 338 subjects, 240 of whom presented symptoms of arterial hypertension. Among these, 140 had controlled hypertension, 78 had uncontrolled hypertension despite treatment and 21 were non-hypertensive.
In total, the dataset comprises 27,586 frames, corresponding to an average of 47.6 frames per video.
All videos are of good quality, with anatomical structures mostly visible, and were manually annotated by an expert. 

\subsubsection{Target Domain}
The target domain $\mathcal{D}_T$ is an in-house, unlabeled, heterogeneous dataset containing 31,053 videos including A4C and A2C views. The videos were acquired from 357 out-patient centers across 22 states in the United States as part of routine clinical care, with approval from the respective ethics committees. 
They were acquired from 15,998 individual subjects, totaling 1,244,844 frames (average of 40 frames per video). The subjects present a broad spectrum of cardiac pathologies, with varying degrees of severity ranging from trace to severe, summarized in Fig.\ref{fig:icardio_data}.

128 full length videos, each from a unique subject, were annotated and validated by an expert to form a test set for evaluating all methods. 
This subset contains 5,175 frames corresponding to an average of 40.4 frames per video.

A common pixel spacing of 0.37 mm was selected based on the average spacing computed over a subset of 1,000 videos from the target domain and was used throughout training.

\subsection{Experimental Setup}
\begin{table*}[!ht] 
\caption{State-of-the-art unsupervised domain adaptation methods and foundation models compared to RL4Seg3D. Best and second best results are respectively emphasized with bold and underlined font. `\textsuperscript{ns}' indicates \underline{no} statistically significant difference from the best method (paired one-tailed t-test, $p \geq 0.05$). Intra-expert variability on the CAMUS dataset is included as a reference upper bound for achievable segmentation quality. ENDO: endocardium, EPI: epicardium, MVC: mitral valve commissure, MpC: mistakes per cycle.}
    \centering
    \resizebox{\linewidth}{!}{
    \begin{tabular*}{\textwidth}{l @{\extracolsep{\fill}} ccccccccc}
    \toprule
    \multirow{2}{*}[-3pt]{Method} & \multicolumn{3}{c}{Dice (\%) $\uparrow$} & \multicolumn{3}{c}{Hausdorff (mm)  $\downarrow$}   & \multirow{2}{*}[-3pt]{\makecell{Anatomical \\ Validity (\%) $\uparrow$}} &  \multirow{2}{*}[-3pt]{\makecell{Temporal \\ Validity (\%) $\uparrow$}} & \multirow{2}{*}[-3pt]{\makecell{MVC Landmark \\ 7.5mm MpC $\downarrow$}} \\
     \cmidrule(lr){2-4}  \cmidrule(lr){5-7}
    & \footnotesize{ENDO.} &  \footnotesize{EPI.} & \footnotesize{Avg.} & \footnotesize{ENDO.} &  \footnotesize{EPI.} & \footnotesize{Avg.}  &  &  & \\

\midrule\midrule

Intra-expert var. (\footnotesize{CAMUS \cite{leclerc2019deep}})~ & 94.4 & 95.4 & 94.9 & \phantom{*}4.3 & \phantom{*}5.0 & \phantom{*}4.6 & 100 & - & \phantom{*}- \\

\midrule

Baseline 3D U-Net & 90.8 & 93.6 & 92.2 & 8.5 & \phantom{*}9.4 & \phantom{*}8.9 & 27.3  & 23.4 & \phantom{*}5.4 \\
nnU-Net \cite{Isensee2021-nnunet} & \phantom{*}\underline{92.6}\textsuperscript{ns} & \phantom{*}95.0\textsuperscript{ns} & \phantom{*}93.8\textsuperscript{ns} & 6.2 & \phantom{*}9.5 & \phantom{*}7.8 & 48.4 & 46.9 & \phantom{*}\textbf{0.6} \\
MedSAM \cite{ma2024segment} & 90.5 & 81.1 & 85.8 & 6.3 & 14.3 & 10.3 & \phantom{0}0.0 & \phantom{0}0.0 & 34.7 \\
SAMUS \cite{lin2024samus} & 88.3 & 93.7 & 91.0 & 7.2 & 15.9 & 11.6 & \phantom{0}0.0 & \phantom{0}0.0 & \phantom{*}5.5\\
MemSAM \cite{deng2024memsam} & 90.0 & 93.2 & 91.6 & 7.4 & \phantom{*}8.1 & \phantom{*}7.7 & 48.4 & 39.8 & \phantom{*}6.0 \\
MaskedSSL \cite{maani2024simlvseg} & 91.5 & 95.1 &  93.3 & 6.2 & \phantom{*}6.4 & \phantom{*}6.3 & 64.1 & 56.3 & \phantom{*}3.1 \\
UA-MT \cite{yu2019uncertainty} & 91.5 & 94.4 &  93.0 & 6.8 & \phantom{*}8.1 & \phantom{*}7.4 & 31.3 & 26.6 & \phantom{*}4.5 \\
\midrule
RL4Seg (2D) \cite{Judge_RL4Seg} & 91.5 & 94.9 & 93.2 & 5.6 & \phantom{*}5.8 & \phantom{*}5.7 & 84.4 & 58.6 & \phantom{*}2.5 \\
\colrow
RL4Seg3D \footnotesize{(Anat. only)} & 92.2 & 94.7 & 93.5 & 5.3 & \phantom{*}6.1 & \phantom{*}5.7 & \phantom{*}\underline{98.4}\textsuperscript{ns} & 78.1 & \phantom{*}4.6 \\
\colrow
RL4Seg3D \footnotesize{(Anat.+LM)} & \textbf{92.8} & \textbf{95.6} & \textbf{94.2} & \phantom{*}\underline{4.9}\textsuperscript{ns} & \phantom{**}\underline{4.9}\textsuperscript{ns} & \phantom{**}\underline{4.9}\textsuperscript{ns} & \phantom{}96.9 & \phantom{}{85.9} & \phantom{**}1.1\textsuperscript{ns} \\
\colrow
RL4Seg3D \footnotesize{(Anat.+LM+T.Pen.)} & 92.5 & \underline{95.4} & \underline{94.0} & \phantom{*}5.0\textsuperscript{ns} & \phantom{**}5.1\textsuperscript{ns} & \phantom{**}5.0\textsuperscript{ns} & \phantom{*}{97.7}\textsuperscript{ns} & \phantom{}\underline{88.3} & \phantom{**}1.2\textsuperscript{ns} \\
\colrow
RL4Seg3D \footnotesize{(Test-time optim.)} & \textbf{92.8} & \textbf{95.6} & \textbf{94.2} & \textbf{4.7} & \phantom{*}\textbf{4.8} & \phantom{*}\textbf{4.7} & \textbf{99.2} & \textbf{93.0} & \phantom{**}\underline{1.0}\textsuperscript{ns} \\

    \bottomrule   
    \end{tabular*}
    }
\label{tab:results}
\vspace{-10pt}
\end{table*}

\subsubsection{Baseline and Comparative Methods}
To establish a baseline for segmentation performance on the target domain, we used the standard 3D U-Net \cite{ronneberger2015unet} and nnU-Net \cite{Isensee2021-nnunet}. Both models were trained exclusively on the source domain and evaluated on the target domain without any form of adaptation.

\textbf{\em Foundation Models:}
We tested several SAM-based foundation models including, MedSAM \cite{ma2024segment} and SAMUS \cite{lin2024samus}, using their provided pre-trained weights. Also, we evaluated MemSAM \cite{deng2024memsam} after fine-tuning on the CAMUS \cite{leclerc2019deep} dataset with full-frame annotations, as recommended by the authors.

\textbf{\em Unsupervised DA Methods:}
The unsupervised methods we evaluated include a masked self-supervised learning scheme inspired by SimLVSeg \cite{maani2024simlvseg}, where a 3D U-Net is pre-trained to reconstruct target domain image sequences from randomly masked inputs, followed by supervised training on the source domain. We also included an uncertainty-aware mean teacher (UA-MT) approach \cite{yu2019uncertainty}, in which a teacher model provides uncertainty-guided feedback to a student model on the target domain. Lastly, the original 2D version of RL4Seg \cite{Judge_RL4Seg} was trained using extracted end-diastole and end-systole frames from the target domain data, and evaluated on all frames of the test videos.

\subsubsection{RL4Seg3D Training Configuration}
RL4Seg3D models were trained using approximately 32 NVIDIA A100 GPUs, though memory requirements did not necessitate full use of the 40 GB capacity. End-to-end training required approximately 2 days, consisting of 2–3 iterations of the loop with anatomical reward network training (50 epochs) followed by 10 epochs of RL optimization, with the target dataset size doubled at each iteration. The number of GPUs used can vary according to resource availability and requirements related to dataset size, influencing training time and effective batch size (see Sec.\ref{sec:2dt_sequences}).
The divergence constraint and entropy coefficients used were respectively $\beta\!=\!0.015$ and $\alpha\!=\!0.15$.
Model selection for final evaluation across training epochs and RL loop iterations was based on the highest validation reward.

\subsubsection{Evaluation}
\label{sec:eval_metrics}
We evaluate all methods based on multiple key segmentation criteria. Post-processing was applied across all methods to remove disconnected regions and ensure optimal results. First, overall segmentation quality, with Dice coefficient and Hausdorff distance, is reported for the endocardium (ENDO), epicardium (EPI) and their average. As these metrics alone cannot fully represent clinically relevant segmentation performance, exclusive reliance on them could lead to misleading comparisons and reduce clinical applicability \cite{maier2024metrics}.

For this reason, we include specific echocardiography metrics. Anatomical validity is defined according to 10 criteria\footnote{Presence of Left ventricle (LV) and myocardium (MYO), holes in LV and MYO, LV and MYO disconnectivity, holes between LV and MYO, LV and BG frontier ratio, MYO thickness, LV width to MYO thickness ratio.}
\cite{Painchaud_2020}, and a segmentation is only valid if all criteria are met across all frames. We also consider temporal validity based on the smoothness of temporal evolution of 8 segmentation attributes\footnote{LV and MYO area, EPI center of mass (X and Y), LV length, LV base width, Hausdorff distance between neighboring frames (MYO and EPI).} \cite{Painchaud_2022}. Each temporal attribute is computed independently for all frames, and a frame is flagged as inconsistent if its value deviates from the linear interpolation of its neighboring frames by more than an attribute-specific threshold. A sequence is considered temporally consistent if no frames are flagged across any attributes. For both anatomical and temporal consistency, we report overall sequence validity, averaged across the test set.

We evaluate mitral valve commissure (MVC) landmark precision using the ``mistakes per cycle (MpC)'' metric, which counts instances where any of the predicted segmentation's landmarks points are further than 7.5 mm from the ground truth's reference landmark locations within a sequence, normalized by the number of cardiac cycles. As the length of the commissural tissue varies between 5 to 10 mm \cite{dal2013anatomy}, 7.5 mm mistakes indicate substantial localization errors, where the segmentation does not align with the underlying anatomy. Valve location is extracted automatically at the junction of the left-ventricle, myocardium and background classes.

\subsection{Results}
Table \ref{tab:results} presents results on the expert validated test set from the target domain. The nnU-Net offers a strong baseline, outperforming the classic 3D U-Net, and yielding very high mitral valve commissure precision. Foundation models, most notably MemSAM, show competitive results, but are affected by spatial and temporal irregularities, leading to high Hausdorff distances as well as poor anatomical and temporal validity. Unsupervised methods show strong performance, consistently achieving high Dice and Hausdorff scores. Among them, the RL based methods most effectively address the domain adaptation challenges, leading to the lowest Hausdorff distances and high anatomical validity. RL4Seg3D achieves the best performance, with particularly strong anatomical and temporal validity, and precise mitral valve commissure localization.

\begin{figure*}[tp]
    \centering \includegraphics[width=1\linewidth]{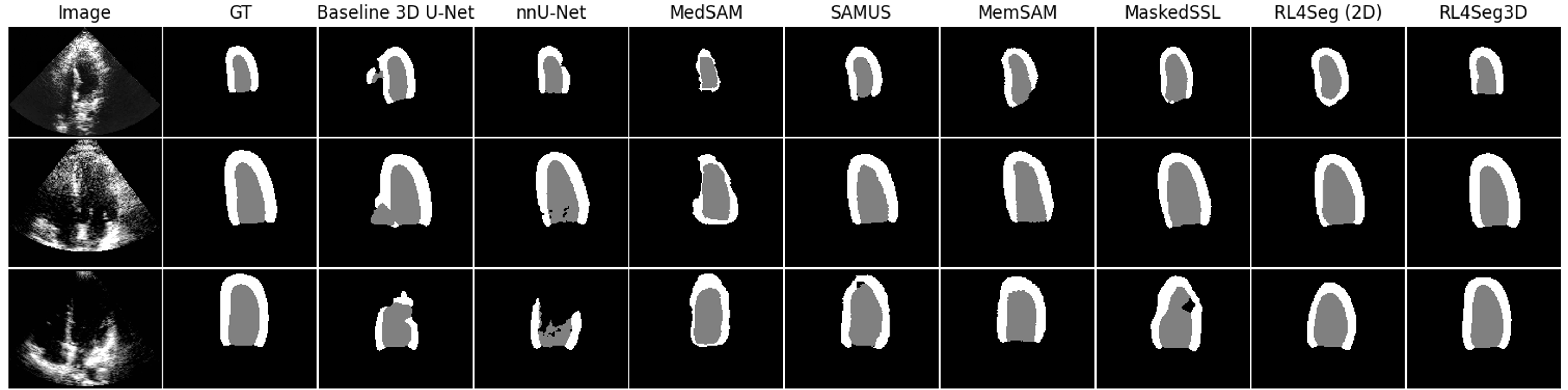}
    \caption{Qualitative comparison of segmentations. Displayed frames
    correspond to the lowest Dice score between the ground truth and baseline 3D U-Net in their respective videos. Full sequence comparisons are provided in the supplementary material.}
    \label{fig:qualitative_comparison}
\end{figure*}

Qualitative results are shown in Fig.\ref{fig:qualitative_comparison} for challenging frames in various sequences. The trends observed in table~\ref{tab:results} are reflected in the segmentation maps, where most methods exhibit spatial irregularities and temporal inconsistencies (sup. mat.), often resulting in anatomically implausible structures. In contrast, RL4Seg3D produces anatomically coherent segmentations, demonstrating robustness across entire echocardiographic sequences, most particularly in the challenging frames, where some structures are not entirely visible.

\subsubsection{Temporal Consistency}

\begin{figure}[ht]
    \centering
    \includegraphics[width=\linewidth]{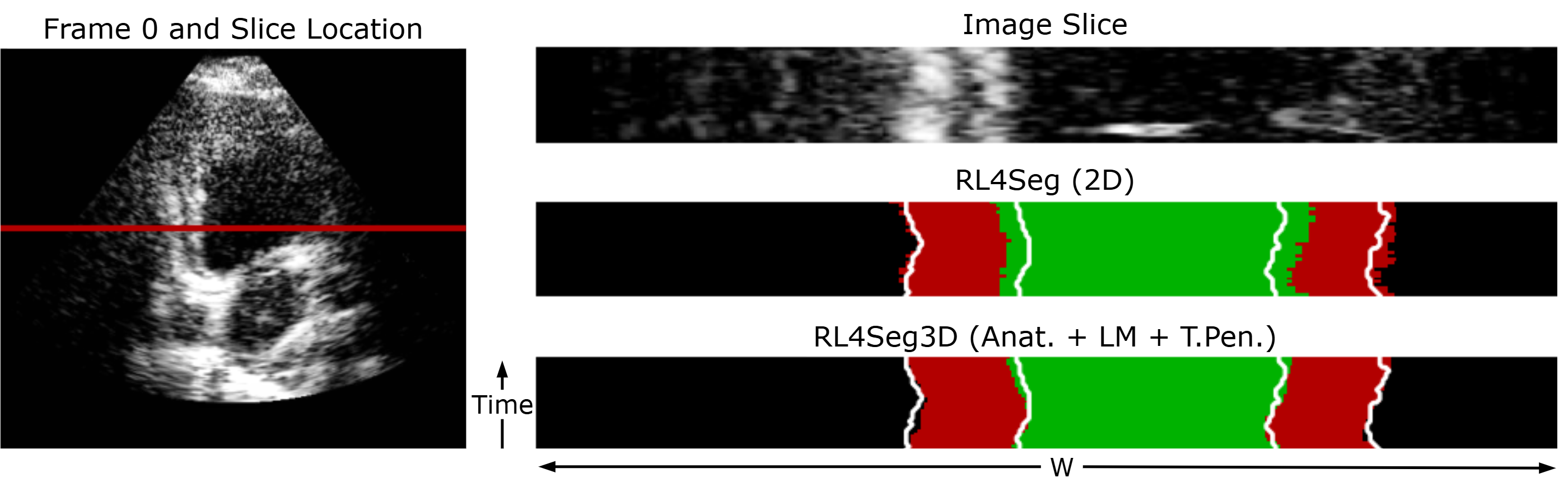}
    \caption{Comparison of the temporal evolution of a slice from segmentations from RL4Seg (2D) and RL4Seg3D, compared to the ground truth %segmentation 
    (white outline). Protrusions show inconsistent temporal evolution of the segmentation between frames. The example shown is representative of the test set, with metrics within one standard deviation of the mean.
    % \vspace{-0.5cm}
    }
    \label{fig:temporal_consistency}
\end{figure}

Table \ref{tab:results} shows that RL4Seg3D produces segmentations with a significantly higher rate of temporal consistency compared to other methods, particularly those relying on 2D processing. Figure \ref{fig:temporal_consistency} presents temporal evolution of a slice from segmentations obtained from the original 2D RL4Seg and RL4Seg3D with the ground truth outline. RL4Seg3D not only better aligns with the ground truth, it offers much smoother temporal dynamics. The 2D method's many irregularities and protrusions reflect temporal inconsistency, which manifest as oscillations or shaking of the segmentation boundary when viewed as a video sequence.
While Fig.\ref{fig:temporal_consistency} illustrates these dynamics on a representative test sequence, across the entire test set, the 2D method produces on average 2.7 temporally inconsistent frames per sequence (average sequence length in the test set is 40.4 frames), compared to only 0.4 for RL4Seg3D (0.2 with TTO).

\subsubsection{Reward Integration}

Table \ref{tab:results} also illustrates the impact of each reward component in the RL framework. Incorporating the landmark reward network significantly enhances mitral valve commissure localization, as reflected by a reduction in 7.5 mm mistakes per cycle. Both the Dice score and Hausdorff distance (HD) also improve, suggesting better alignment with the ground truth segmentation. The slight decrease in anatomical validity reflects a trade-off between precise structural alignment and anatomical plausibility. Results also highlight the impact of integrating the temporal penalty into the fused rewards. By penalizing frames with temporal inconsistencies, the policy learns a smoothness prior, improving temporal validity with minimal effect on other metrics. 

These results underscore the flexibility of the RL framework, which allows for tuning and balancing various reward signals to target specific aspects of segmentation.

\subsubsection{Uncertainty Estimation}

\begin{figure*}[tp]
    \centering \includegraphics[width=\linewidth]{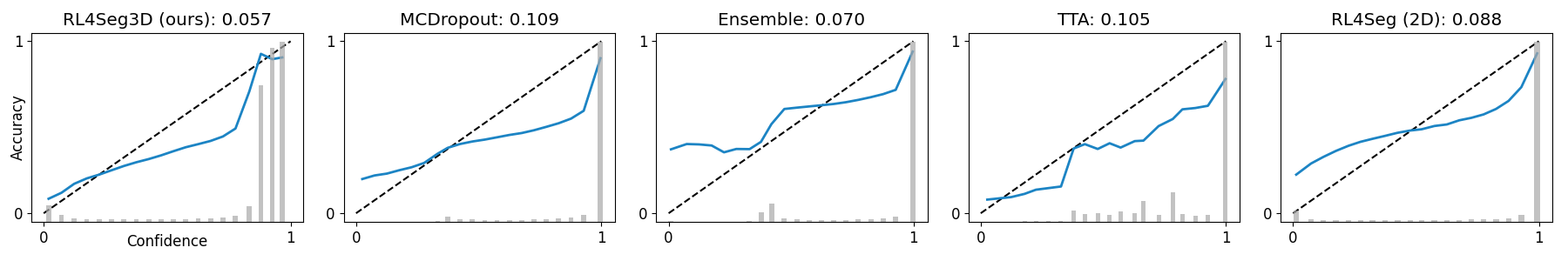}
    \caption{Calibration plots for uncertainty methods, with expected calibration error (ECE) (the smaller the better). Grey histograms in the background represent the density of pixels in each confidence bin.}
    \label{fig:ECE}
\end{figure*}

As mentioned in Sec.\ref{sec:post-proc-val}, RL4Seg3D's anatomical reward network can act as an effective uncertainty estimator. We compare the uncertainty estimates provided by the temperature-scaled 3D anatomical reward network to its 2D counterpart, as well as to established epistemic uncertainty estimation methods such as Monte-Carlo Dropout (MCDropout) \cite{gal2016dropout} and model ensembling \cite{NIPS2017_ensembleUNC}. We also include test-time augmentation \cite{wang2019aleatoricTTA}, an aleatoric uncertainty estimator. To quantify uncertainty quality, we compute the expected calibration error (ECE) between predicted uncertainty maps and ground truth error maps. Figure~\ref{fig:ECE} demonstrates the superior calibration performance of the 3D anatomical reward network $r_{\psi}^{\mathit{ANAT}}$, as it provides better-aligned uncertainty estimates with observed segmentation errors. For calibration evaluation, which assesses pixel-level errors against predicted confidence, we only use the anatomical reward network as the landmark reward network produces coarser error regions, which reduces calibration (with reward fusion, $\textup{ECE} = 0.067$).

\begin{figure}[ht]
    \centering
    \begin{subfigure}[t]{0.32\linewidth}
        \centering
        \includegraphics[width=\linewidth]{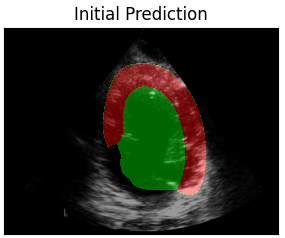}
        \caption{}
        \label{subfig:TTO_initial}
    \end{subfigure}
    \begin{subfigure}[t]{0.32\linewidth}
        \centering
        \includegraphics[width=\linewidth]{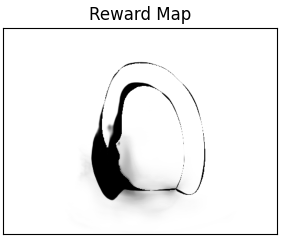}
        \caption{}
        \label{subfig:TTO_reward}
    \end{subfigure}
    \centering
    \begin{subfigure}[t]{0.32\linewidth}
        \centering
        \includegraphics[width=\linewidth]{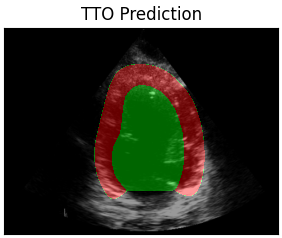}
        \caption{}
        \label{subfig:TTO_final}
    \end{subfigure}
    \caption{(a) Initial segmentation containing an invalid anatomical shape due to poor signal on the septal wall. (b) Corresponding reward map highlighting the error. (c) Final segmentation after test-time-optimization (TTO), with the error corrected. The frame shown has the lowest reward in the sequence; the full video is provided in the supplementary material. %\vspace{-0.5cm}
    }
    \label{fig:TTO}
\end{figure}

\subsubsection{Test-Time Optimization}
We also demonstrate the effectiveness of the test-time optimization (TTO) scheme, guided by the high-quality uncertainty maps. Table \ref{tab:results} shows TTO improves segmentation for videos where the policy initially produced either anatomically or temporally invalid outputs (22 of the 128 subjects). %The pixel-wise reward maps allow the policy to adapt to specific videos by focusing optimization on problematic frames, while frames with already high confidence and accuracy (therefore high reward) remain largely unchanged. 
As can be seen, errors are corrected, leading to improvements across all metrics.  Since only a minority of the videos have been updated, global scores like the Dice remain relatively unchanged, while other metrics got an important boost, especially the temporal validity which went from 88.3\% to 93\%.

Figure \ref{fig:TTO} illustrates the results of TTO on a specific example containing certain anatomically invalid frames. The reward highlights the error regions and a small number of optimization steps allows the policy to adapt to the specific, most challenging part of the image sequence, which contains errors. Several videos illustrating the effect of TTO are provided in supplementary materials.

\subsubsection{Ablation Studies and Extended Evaluation}
We evaluate the impact of various components of the framework on performance through ablation and sensitivity studies. Due to the length and computational demands of training on the full dataset, for these experiments, we restrict data to a subset of 5000 videos, using fixed reward networks for 10 epochs of RL training.
We report Dice score and Hausdorff distance averaged over the endocardium and epicardium, anatomical validity (AV) and temporal validity (TV) as the percentage of fully valid sequences, and mitral valve commissure landmark mistakes per cardiac cycle, measured with a 7.5mm tolerance.

\textbf{\em Reward Fusion Mechanisms:} We conducted an ablation study on various reward fusion operators, including the minimum, mean, generalized mean and weighted sums of reward maps. For generalized mean, we considered the harmonic mean ($p=-1$) and a parametrization approaching the minimum ($p=-5$), where $p$ is the exponent that controls the type of averaging. Table~\ref{tab:ablation_reward_fusion} reports results obtained with each operator and shows limited variability on the final performance relative to the initial starting point.

\begin{table}[h]
    \caption{Ablation study of reward fusion operators.}
    \centering
    \footnotesize
    \setlength{\tabcolsep}{4.75pt} % tighter column spacing
    \resizebox{\linewidth}{!}{
    \begin{tabularx}{\linewidth}{l c c c c c c}
    \toprule
    \makecell{Fusion \\ Operator} &
    \makecell{Dice \\ (\%) $\uparrow$} &
    \makecell{HD \\ (mm) $\downarrow$} &
    \makecell{AV \\ (\%) $\uparrow$} &
    \makecell{TV \\ (\%) $\uparrow$} &
    \makecell{MVC LM \\ 7.5mm MpC \\ $\downarrow$} \\
    \midrule\midrule
    Baseline (no RL)        & 92.2 & 8.9 & 28.1 & 23.4 & 5.4 \\
    Min.                    & 94.0 & 5.3 & 87.1 & 66.8 & 2.3 \\
    Mean                    & 94.1 & 5.2 & 88.3 & 80.1 & 2.2 \\
    Gen. Mean: $p=-1$       & 93.9 & 5.7 & 82.4 & 66.4 & 2.5 \\
    Gen. Mean: $p=-5$       & 94.0 & 5.2 & 88.3 & 76.2 & 1.9 \\
    Weighted Sum $0.7/0.3$  & 93.9 & 5.4 & 87.1 & 75.4 & 1.7 \\
    Weighted Sum $0.3/0.7$  & 94.1 & 5.1 & 82.8 & 77.3 & 1.9 \\
    \bottomrule
    \end{tabularx}
    }
    \label{tab:ablation_reward_fusion}
\end{table}

All combinations of reward maps contain the same underlying information with slightly varied weightings, leading to similar outcomes across the evaluated fusion operators, as the resulting learning signals remain comparable and PPO updates are dominated by the highest advantage signals. We select the minimum operator for consistency and simplicity, as it preserves the maximum penalty from each reward without introducing any additional hyperparameters.

\textbf{\em Sliding Window Length:} We examined different choices for sliding window length in the framework, in the context of echocardiography. Multiple baseline models were trained on the source domain with windows of lengths 4, 8, 16 and 24 frames. Performance, evaluated on the source domain to assess initial segmentation quality, and RL training computational requirements are summarized in table~\ref{tab:SW_ablation}.

\begin{table}[h]
    \caption{Ablation study of sliding window length. Supervised training results reported on the source domain. Maximum memory (GB) and training time (iter/s) for RL training results are computed from a sample of the target domain.}
    \centering
    \footnotesize
    \setlength{\tabcolsep}{3pt} % tighter column spacing
    \resizebox{\linewidth}{!}{
    \begin{tabularx}{\linewidth}{l c c c c c c c c}
    \toprule
    \multicolumn{1}{c}{} &
    \multicolumn{5}{c}{\textbf{Supervised Training} ($\mathcal{D}_S$)} & 
    \multicolumn{2}{c}{\textbf{RL Training} ($\mathcal{D}_T$)}\\
    \cmidrule(lr){2-6}
    \cmidrule(lr){7-8}
    \makecell{Window \\ Length} &
    \makecell{Dice \\ (\%) $\uparrow$} &
    \makecell{HD \\ (mm) $\downarrow$} &
    \makecell{AV \\ (\%) $\uparrow$} &
    \makecell{TV \\ (\%) $\uparrow$} &
    \makecell{MVC LM \\ 7.5mm MpC \\ $\downarrow$} &
    \makecell{Mem. \\ (GB) $\downarrow$} &
    \makecell{Iter/s $\uparrow$} \\
    \midrule\midrule
    4  & 95.4 & 4.4 & 81.9 & 84.5 & 2.0 & 12 & 0.90 \\
    8  & 95.3 & 4.7 & 74.1 & 81.9 & 2.0 & 25 & 0.48 \\
    16 & 95.0 & 4.8 & 78.5 & 85.4 & 2.2 & 52 & 0.20 \\
    24 & 94.7 & 4.8 & 86.2 & 93.1 & 2.3 & 76 & 0.16 \\
    \bottomrule
    \end{tabularx}
    }
    \label{tab:SW_ablation}
\end{table}

\begin{figure}[ht]
    \centering
    \begin{subfigure}[t]{0.45\linewidth}
        \centering
        \includegraphics[width=\linewidth]{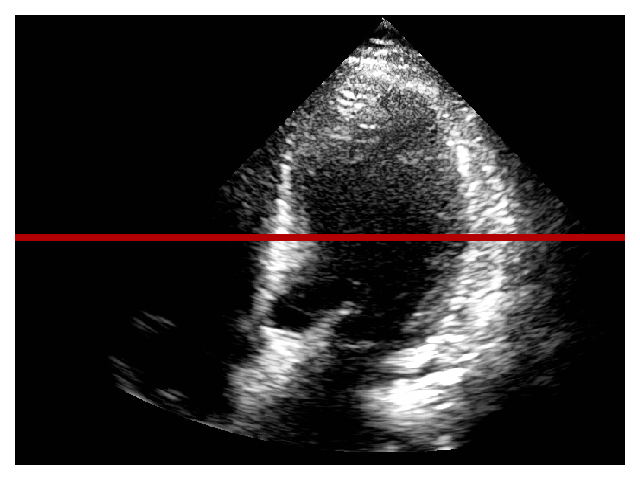}
        \caption{Slice Location}
        \label{subfig:SW_ablation}
    \end{subfigure}
    \begin{subfigure}[t]{0.45\linewidth}
        \centering
        \includegraphics[width=\linewidth]{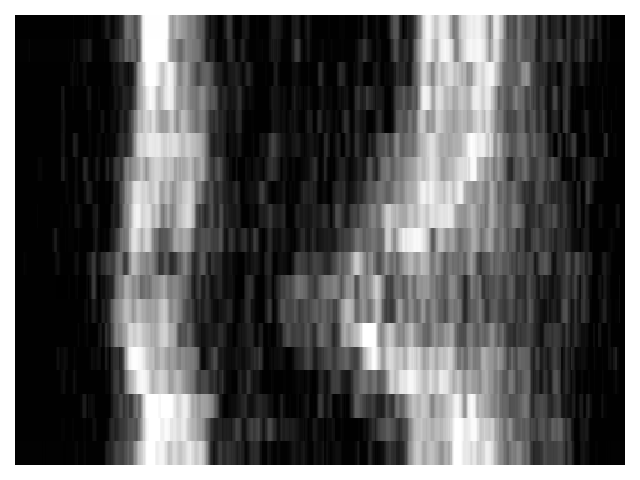}
        \caption{Image Slice}
        \label{subfig:SW_ablation}
    \end{subfigure}
    \begin{subfigure}[t]{0.45\linewidth}
        \centering
        \includegraphics[width=\linewidth]{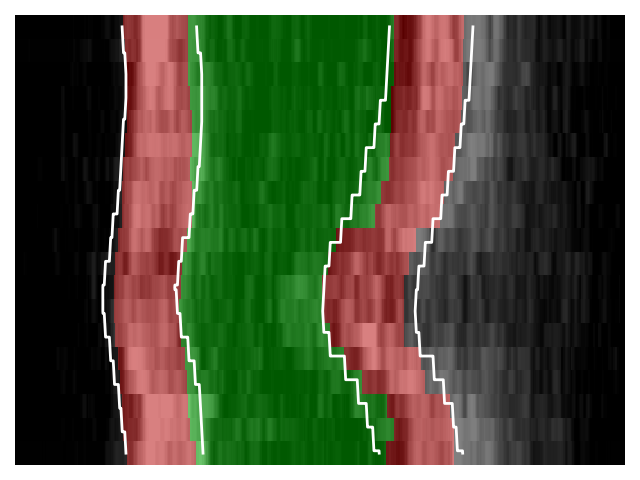}
        \caption{4 Frame Window}
        \label{subfig:SW_ablation}
    \end{subfigure}
    \begin{subfigure}[t]{0.45\linewidth}
        \centering
        \includegraphics[width=\linewidth]{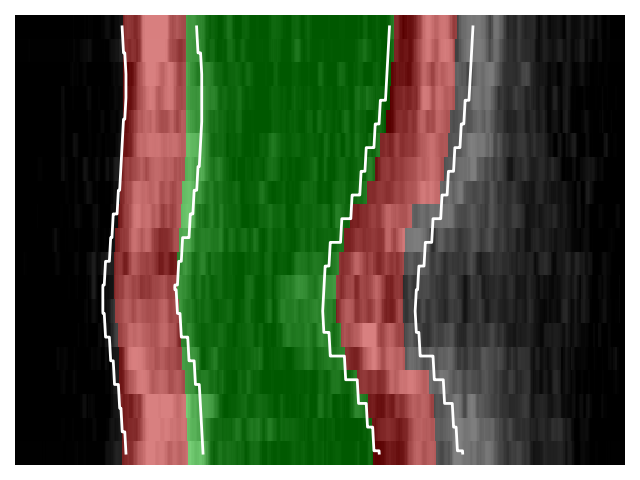}
        \caption{8 Frame Window}
        \label{subfig:SW_ablation}
    \end{subfigure}
    \begin{subfigure}[t]{0.45\linewidth}
        \centering
        \includegraphics[width=\linewidth]{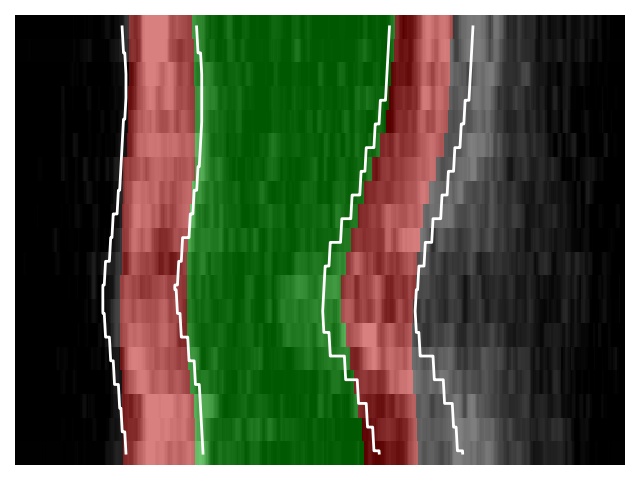}
        \caption{16 Frame Window}
        \label{subfig:SW_ablation}
    \end{subfigure}
    \begin{subfigure}[t]{0.45\linewidth}
        \centering
        \includegraphics[width=\linewidth]{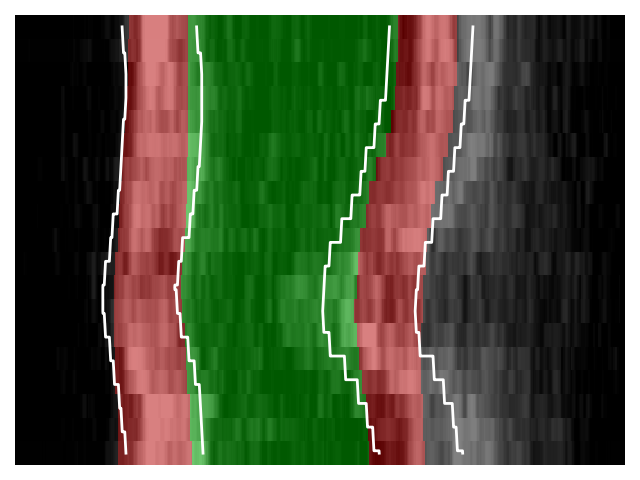}
        \caption{24 Frame Window}
        \label{subfig:SW_ablation}
    \end{subfigure}
    \caption{Comparison of different sliding window length baseline models. Reference segmentation outline shown in white. (b)-(f) show the temporal evolution (vertical axis) along the slice indicated by a red horizontal line in (a).}
    \label{fig:SW_ablation}
\end{figure}

Longer sliding windows benefit from increased temporal context and stronger regularization during training, yielding higher anatomical and temporal validity. However, other metrics show no substantial gains for these models. While temporal validity improves with longer windows, over-smoothing becomes increasingly problematic, a notable issue in echocardiography, where rapid cardiac motion is common although not found in all acquisitions. Figure~\ref{fig:SW_ablation} illustrates that long-window models fail to capture fast movements. 

Jitters and anatomical errors, more common in short-windowed models, can be mitigated with our RL framework and reward mechanisms, whereas slow-moving long sliding windows cannot be made to respond faster. In addition, these models incur a large computational burden. Training time and memory requirements in table~\ref{tab:SW_ablation} demonstrate the prohibitively high costs for long sliding windows. Therefore, a short sliding window of four frames is a balanced and practical choice for echocardiography segmentation.

\textbf{\em Hyperparameter Sensitivity:} We also evaluated the effect of the entropy weight and divergence coefficient, from eq.~\ref{eq:full_loss} and eq.~\ref{eq:div_constraint}, respectively, on performance. Table~\ref{tab:Hyperparam_sensitivity} reports results from a grid search over different hyperparameters settings.

\begin{table}[h]
    \caption{Hyperparameter sensitivity study for different loss terms in the framework.}
    \centering
    \footnotesize
    \setlength{\tabcolsep}{4pt} % tighter column spacing
    \resizebox{\linewidth}{!}{
    \begin{tabularx}{\linewidth}{c c c c c c c c}
    \toprule
    \makecell{Entropy \\ Weight $\alpha$} &
    \makecell{Divergence \\ Coefficient $\beta$} &
    \makecell{Dice \\ (\%) $\uparrow$} &
    \makecell{HD \\ (mm) $\downarrow$} &
    \makecell{AV \\ (\%) $\uparrow$} &
    \makecell{TV \\ (\%) $\uparrow$} &
    \makecell{MVC LM \\ 7.5mm MpC \\ $\downarrow$} \\
    \midrule\midrule
    \multirow{3}{*}{0.30} & 0.001 & 93.8 & 10.2 & \phantom{*}0.0 & \phantom{*}1.6 & 10.5 \\
                          & 0.015 & 94.1 & \phantom{*}5.1 & 77.3 & 71.9 & \phantom{*}1.9 \\ 
                          & 0.100 & 93.8 & \phantom{*}6.2 & 66.4 & 51.6 & \phantom{*}3.3 \\ \hline
    \multirow{3}{*}{0.15} & 0.001 & 93.5 & \phantom{*}5.5 & 91.4 & 72.7 & \phantom{*}3.3 \\
                          & 0.015 & 94.0 & \phantom{*}5.2 & 88.3 & 80.5 & \phantom{*}2.1 \\
                          & 0.100 & 93.9 & \phantom{*}6.0 & 75.8 & 50.0 & \phantom{*}2.7 \\\hline
    \multirow{3}{*}{0.01} & 0.001 & 94.2 & \phantom{*}5.2 & 75.8 & 67.2 & \phantom{*}1.7 \\
                          & 0.015 & 94.3 & \phantom{*}5.3 & 73.4 & 63.3 & \phantom{*}2.2 \\
                          & 0.100 & 93.5 & \phantom{*}5.5 & 91.4 & 72.7 & \phantom{*}3.3 \\
    \bottomrule
    \end{tabularx}
    }
    \label{tab:Hyperparam_sensitivity}
\end{table}

Results show that both entropy and divergence have a large impact on training results and stability, and must be tuned jointly. Their interaction dictates the balance between exploration and optimization stability. While many combinations yield competitive results, $\alpha=0.15$ and $\beta=0.015$ provide the best trade-off, offering the most consistent performance across all metrics.

\textbf{\em Temporal Penalty Scaling:}
To understand the effects of the temporal penalty on training stability and performance, we vary its scaling factor $\kappa$, regulating how strongly inconsistencies are penalized though the Gaussian kernel $G_{\sigma(\eta_t, \kappa)}$. As shown in table~\ref{tab:temp_pen_sensitivity}, increasing the scaling factor yields higher temporal validity, indicating that the penalty has the desired effect. Other metrics remain generally stable, demonstrating the robustness of the framework to heavily distorted reward maps at high scaling factors.

\begin{table}[h]
    \caption{Temporal penalty scaling factor ($\kappa$) sensitivity study.}
    \centering
    \footnotesize
    \resizebox{\linewidth}{!}{
    \begin{tabularx}{\linewidth}{c c c c c c c}
    \toprule
    \makecell{$\kappa$ \\ Scaling Factor} &
    \makecell{Dice \\ (\%) $\uparrow$} &
    \makecell{HD \\ (mm) $\downarrow$} &
    \makecell{AV \\ (\%) $\uparrow$} &
    \makecell{TV \\ (\%) $\uparrow$} &
    \makecell{MVC LM \\ 7.5mm MpC $\downarrow$} \\
    \midrule\midrule
    1  & 93.9 & 5.3 & 82.8 & 66.4 & 2.8 \\
    10 & 94.1 & 5.2 & 91.4 & 71.9 & 1.7 \\
    50 & 93.9 & 5.2 & 86.7 & 75.0 & 1.8 \\
    \bottomrule
    \end{tabularx}
    }
    \label{tab:temp_pen_sensitivity}
\end{table}

However, an important caveat is that at lower effective batch sizes, training stability is compromised at high scaling factors. The reduced regularization causes highly penalized reward maps to exert a disproportionately large influence on the policy, ultimately derailing training. A strong temporal penalty requires a sufficiently large batch size to maintain stable optimization.

\subsubsection{Computational Efficiency}
In order to further compare RL4Seg3D to other key state-of-the-art methods, we conducted a comparison of computational efficiency during training and inference. Table~\ref{tab:compt_requirements} reports these results, showing that RL4Seg3D is competitive remains competitive across both settings.

\begin{table}[h]
    \caption{Computational requirements and resource usage comparison of key methods for both training and inference.}
    \centering
    \footnotesize
    \setlength{\tabcolsep}{2pt} % tighter column spacing
    \begin{tabularx}{\linewidth}{l c c c c c c}
        \toprule
        \multicolumn{1}{c}{} &
        \multicolumn{2}{c}{\textbf{Training}} & 
        \multicolumn{4}{c}{\textbf{Inference}}\\
        \cmidrule(lr){2-3}
        \cmidrule(lr){4-7}
        \makecell{Method} &
        \makecell{Iter/s $\uparrow$} &
        \makecell{Mem. \\ (GB)  $\downarrow$} &
        \makecell{Input Size \\ (pixels)} &
        \makecell{FLOPs $\downarrow$} &
        \makecell{Time \\(s)  $\downarrow$} &
        \makecell{Mem. \\(GB) $\downarrow$} \\
        \midrule\midrule
        Baseline (SFT) & 4.2 & 12.2 & \phantom{**}736x544 & $3.5\times10^{7}$ & \phantom{*}1.3 & \phantom{*}3.1 \\
        nnU-Net        & 1.3 & \phantom{*}8.0 & \phantom{**}736x544 & $7.9\times10^{7}$ & \phantom{*}1.6 & \phantom{*}2.6 \\
        MemSAM         & 1.3 & 17.9 & 3x256x256 & \phantom{*}$1.2\times10^{13}$ & \phantom{*}1.1 & \phantom{*}7.0 \\
        UA-MT          & 1.9 & 15.6 & \phantom{**}736x544 & $3.5\times10^{7}$ & \phantom{*}1.3 & \phantom{*}3.1\\
        RL4Seg3D       & 0.7 & 12.5 & \phantom{**}736x544 & $3.5\times10^{7}$ & \phantom{*}1.3 & \phantom{*}3.1 \\
        TTO (RL4Seg3D) & N/A & N/A & \phantom{**}736x544 & \phantom{*}$1.4\times10^{10}$ & 96.9 & 11.4 \\
        \bottomrule
    \end{tabularx}
    \label{tab:compt_requirements}
\end{table}

Training-wise, RL4Seg3D is moderately more computationally demanding than other methods, requiring more training time. This overhead is primarily due to the multiple optimization steps per mini-batch inherent to the PPO algorithm. Nevertheless, given the performance gains achieved, the method remains practical. In contrast, approaches such as UA-MT and MemSAM incur substantially larger memory usage, due to the dual-network (student-teacher) setup and the large transformer backbone, respectively. Since many components in RL4Seg3D do not require simultaneous gradient computation, its overall memory footprint remains reasonable.

During inference, as the baseline, nnU-Net, UA-MT and RL4Seg3D rely on the same network architecture, computational costs are comparable. MemSAM, however, inherits a transformer-based backbone from SAM, leading to significantly higher computational complexity and memory consumption, even at reduced spatial resolution.

While test-time optimization substantially reduces inference efficiency, it is the only method that performs inference-time backpropagation to refine predictions. As such, TTO is best suited for a limited subset of low-confidence or failure cases in the context of retrospective analysis.

\subsection{Discussion}
Overall, our results demonstrate that RL4Seg3D provides a robust domain adaptation method for spatio-temporal segmentation of full-sized 2D+t images from a large-scale unlabeled target domain. Its performance surpasses state-of-the-art methods and approaches the upper bound for segmentation accuracy on medium to high-quality images, defined by intra-expert variability on the CAMUS dataset. 
Additionally, RL4Seg3D maintains comparable performance on test subjects exhibiting left ventricular shape anomalies (see Fig.~\ref{fig:icardio_data}), indicating robustness to clinically relevant anatomical variations.
Fusing adaptive and static rewards enables concurrent optimization of different objectives and can be leveraged to generate reliable uncertainty estimates.

Through the reward fusion, the RL framework allows the policy to be optimized not only for segmentation quality and validity, but also for specific issues such as landmark precision 
% \added{, while maintaining robustness to the choice of fusion operator}.
While the temporal reward does not yield statistically significant improvements in aggregate performance metrics, it leads to better results, which is expected given that it is active only for temporally invalid segmentations which represent a limited subset of cases. The inclusion of the landmark-based reward, in contrast, yields statistically significant improvements across all experiments in which it is included.
As a result, predicted mitral valve commissure deviates more than 7.5 mm away from the ground truth's commissure locations in only 1.1 frames per cardiac cycle (average frame count per cycle is 35.2 frames). 3D segmentation RL also allows enforcing of any new prior in the target domain, without the need for a differentiable function, and for it to be combined with any number of priors though reward fusion.

The nnU-Net also demonstrated very high mitral valve commissure precision, with 7.5 mm mistakes per cycle on an average of 0.6 frames. Its 3D patch-wise training and processing gives it an advantage for accurately locating and segmenting this region compared to full-image segmentation. However, it struggles in segmenting regions with little or no signal (see Fig.\ref{fig:qualitative_comparison}), leading to anatomical and temporal errors. Such regions are not common in the source domain which it was trained on, highlighting the need for domain adaptation.

Foundation models provided good zero-shot predictions, showcasing their strong generalization, but inconsistencies reduced their effectiveness. Over-reliance on prompts sometimes caused errors and even the segmentation of the wrong structure, leading to strong failure cases and large Hausdorff distance values. MemSAM's memory reinforcement module mitigated some of these issues, improving performance. MedSAM, however, offered weak myocardium segmentation, reflected by low epicardium scores, likely due to lack of training data, as left-ventricle annotations are more common in datasets used for foundation model training. Ultrasound-specific models (SAMUS, MemSAM) offered much better myocardium segmentation.

Building on this, results highlight the advantages of unsupervised methods that leverage the target domain data. While SAM-based methods benefited from training on large multi-structure, multi-modality datasets, they underperformed compared to SimLVSeg's masked self-supervised learning scheme and the UA-MT mean teacher approach. Their generalization strength comes at the expense of high-precision in specific segmentation tasks. Adapting model weights to the target domain's feature distribution, even via pre-training, improved performance drastically but still lacked the structure level guidance, as provided by segmentation RL reward mechanisms.

By making use of context from neighboring frames, 3D convolutions and temporal constraints (e.g. MemSAM's memory prompting) significantly improved performance across all metrics, most notably temporal validity. This consistency, combined with additional reward components and the processing of full-sized inputs allows RL4Seg3D to significantly outperform its 2D counterpart, without additional labels on the target domain, and makes it suitable for extension to volumetric data.

Beyond a strong segmentation policy, the RL4Seg3D framework provides robust uncertainty estimation that outperforms state-of-the-art uncertainty methods in expected calibration error. 
While this uncertainty guides domain adaptation of the policy and can be used to identify high-confidence segmentations, it can also be exploited to refine the policy on specific challenging videos with test-time optimization (TTO). However, as its substantial computational cost limits TTO's practicality for routine real-time deployment, it is intended for retrospective refinement or targeted application to low-confidence or erroneous cases, where the additional computation is justified to improve predictions. This enables even stronger segmentation performance, with almost all segmentations now temporally and anatomically valid. Further examination reveals that remaining errors correspond mostly to regions of low signal-to-noise ratio, imaging artifacts or cardiac structures such as papillary muscles, as well as minor temporal inconsistencies caused by rapid cardiac motion, which remains difficult to capture. The latter would likely benefit from a more comprehensive temporal reward mechanism.

\section{Conclusion}
We presented RL4Seg3D, an unsupervised domain adaptation framework for 2D + time spatio-temporal echocardiography segmentation. By extending the application of reinforcement learning to full-length sequences, introducing a flexible sliding window approach that supports high-resolution, full-sized inputs, and fusing multiple reward mechanisms, RL4Seg3D outperforms baselines and foundation models across overall segmentation accuracy and echocardiography-specific metrics including anatomical and temporal validity as well as mitral valve commissure landmark precision. 

We demonstrate RL4Seg3D's effectiveness for reliable, scalable, and annotation-free domain adaptation for segmentation in a clinically challenging domain such as echocardiography, using a large dataset of over 30 000 videos.
We further highlight its potential for unsupervised annotation and sequence-specific adaptation via calibrated uncertainty estimates and test-time optimization, which supports robust generalization to new datasets.

\appendices

%\section*{awledgment}

\bibliographystyle{IEEEtran}
\bibliography{references.bib}

\end{document}